\documentclass[fleqn,usenatbib]{mnras}

\usepackage{newtxtext,newtxmath}

\usepackage[T1]{fontenc}

\DeclareRobustCommand{\VAN}[3]{#2}
\let\VANthebibliography\thebibliography
\def\thebibliography{\DeclareRobustCommand{\VAN}[3]{##3}\VANthebibliography}

\usepackage{graphicx}	
\usepackage{amsmath}	
\usepackage{subcaption}
\usepackage{pdflscape}
\usepackage{longtable}
\usepackage{multirow}

\newcommand{\gmrt}{uGMRT}
\newcommand{\vla}{VLA}
\newcommand{\lofar}{LOFAR}
\newcommand{\yarp}{\textsc{YARP}}
\newcommand{\emcee}{\textsc{EMCEE}}
\newcommand{\python}{\texttt{Python}}

\newcommand{\aips}{\textsc{AIPS}}

\newcommand{\fesc}{$f_\mathrm{esc}^\mathrm{LyC}$}
\newcommand{\alphanth}{\ifmmode \alpha_\mathrm{nth} \else $\alpha_\mathrm{nth}$\fi}
\newcommand{\alphacs}{$\alpha^{\mathrm{3GHz}}_\mathrm{6GHz}$}
\newcommand{\Oratio}{O$_{32}$}
\newcommand{\sfrdensity}{$\Sigma_\mathrm{SFR}$}

\newcommand{\hi}{H{\sc i}}
\newcommand{\hii}{H{\sc ii}}
\newcommand{\RCth}{RC$_\mathrm{th}$}

\newcommand{\ewhbeta}{EW(H$\beta$)}
\newcommand{\fth}{\ifmmode \mathrm{f}_{1 \mathrm{GHz}}^\mathrm{th (SED)} \else f$_{1 \mathrm{GHz}}^\mathrm{th (SED)}$\fi}

\newcommand{\sth}{\ifmmode S_{\mathrm{1GHz}}^{\mathrm{th}} \else S$_{\mathrm{1GHz}}^{\mathrm{th}}$\fi}
\newcommand{\snth}{\ifmmode S_{\mathrm{1GHz}}^{\mathrm{nth}} \else S$_{\mathrm{1GHz}}^{\mathrm{nth}}$\fi}
\newcommand{\nuturn}{\ifmmode \nu_\mathrm{t} \else $\nu_\mathrm{t}$\fi}

\newcommand{\odepth}{$\tau_\nu^{\mathrm{ff}}$}

\newcommand{\mstar}{$\mathrm{M_*}/\mathrm{M_\odot}$}
\newcommand{\rcsed}[1]{radio-SED{#1}}
\newcommand{\xsfg}[1]{xSFG{#1}}
\newcommand{\bbten}{J150934+373146}
\newcommand{\bbthirteen}{J160810+352809}
\newcommand{\sbs}{SBS 0335-052}

\newcommand{\ebv}{\textit{E(B-V)}}
\newcommand{\hbeta}{H~$\beta$}
\newcommand{\hbetaradio}{ $\mathrm{Flux}^\mathrm{Exp}_{\mathrm{H}\beta}$  }
\newcommand{\specindex}{$\alpha^{\mathrm{low}}_\mathrm{high}$}
\newcommand{\Lya}{Ly~$\alpha$}
\newcommand{\vsep}{$V_\mathrm{sep}$}
\defcitealias{Hunt04}{H04}
\defcitealias{Sebastian19}{SB19}

\newcommand{\revtext}[1]{#1}
\newcommand{\revtexts}[1]{#1}

\title[Radio-SEDs of metal-poor extreme starbursts]{Radio Spectral Energy Distribution of Low-$z$ Metal Poor Extreme Starburst Galaxies: Novel insights on the escape of ionizing photons}

\author[Bait et al.]{
Omkar Bait,$^{1,2,3}$\thanks{CosmicAI Fellow (obait@nrao.edu (OB))}
Daniel Schaerer,$^{4,5}$
Yuri I. Izotov$^{6}$
and Biny Sebastian$^{7}$
\\
$^{1}$ National Radio Astronomy Observatory, 520 Edgemont Road, Charlottesville, VA 22903, USA \\
$^{2}$ The NSF-Simons AI Institute for Cosmic Origins, USA, 201 E. 24th Street, POB 4.102, Austin, Texas 78712-1229 \\
$^{3}$SKA Observatory, Jodrell Bank, Lower Withington, Macclesfield, SK11 9FT, UK\\
$^{4}$Observatoire de Gen\`eve, Universit\'e de Gen\`eve, Chemin Pegasi, 1290 Versoix, Switzerland\\
$^{5}$CNRS, IRAP, 14 Avenue E. Belin, 31400 Toulouse, France \\
$^{6}$Bogolyubov Institute for Theoretical Physics, 14-b Metrolohichna str, Kyiv 03143, Ukraine \\
$^{7}$Space Telescope Science Institute, 3700 San Martin Drive, Baltimore, MD 21218, USA \\
}

\date{Accepted XXX. Received YYY; in original form ZZZ}

\pubyear{\the\year{}}

\begin{document}
\label{firstpage}
\pagerange{\pageref{firstpage}--\pageref{lastpage}}
\maketitle

\begin{abstract}

\revtexts{Recent optical surveys have identified a rare population of low-$z$ ($z \sim 0.01 - 0.06$) extreme star-forming galaxies (\xsfg{s}) characterized by very low metallicity, strong emission lines, extremely high specific star-formation rate, low stellar mass, and strong Ly~$\alpha$ emission. Their global properties resemble recently discovered $z > 6$ reionization-era star-forming galaxies. We present new multi-frequency radio continuum (RC) observations of $8$ \xsfg{s} using the upgraded Giant Metrewave Radio Telescope (uGMRT) at $1.25$ GHz, the Karl G. Jansky Very Large Array (VLA) at $1.5, 3.0, 6.0, 10.0$ and $15.0$ GHz, along with archival LOw Frequency ARray (LOFAR) data at $150$ MHz for several sources. These data allow construction of the radio spectral energy distribution (\rcsed{}) from $\sim 1$ GHz (down to $150$ MHz for some sources) to $15$ GHz, spanning nearly two orders of magnitude in frequency. We find that \xsfg{s} exhibit a flat spectral index between $6$ and $15$ GHz, while a subset shows spectral turnovers at $0.3 - 3$ GHz. Our Bayesian \rcsed{} modeling indicates that these features are consistent with a high thermal fraction combined with free-free absorption, requiring high emission measures in some systems. By comparing modeled thermal radio emission with observed H$\beta{}$ line flux density, we find evidence for dust in several \xsfg{s}. Finally, we confirm a previously reported correlation between Lyman continuum escape fraction (\fesc{}), ionization state, and radio spectral index, particularly among strong leakers.}

\end{abstract}




%

\section{Introduction}
\label{sec: introduction}
Identifying the nature of sources (at redshift $z > 6$) that led to the reionization of the Universe remains a matter of intense debate. Numerous studies, both on the observational and simulation side, suggest that low-mass dwarf starburst galaxies are likely the dominant sources of reionization \citep[e.g.,][]{Ouchi09, Bouwens15, Robertson15, Atek15}. With the advent of the James Webb Space Telescope (JWST), rest-frame optical spectroscopic studies of reionization-era galaxies ($z \sim 6 - 8$) have become possible revealing their key physical properties. It is found that these galaxies generally have low stellar masses ($\log$ \mstar{} $\leq 9$), compact size ($< 1$ kpc), and high specific star-formation rates sSFR of $\sim 100$ Gyr$^{-1}$ \citep{Schaerer22, Rhoads23, Williams23, Endsley23, Fujimoto23_JWST,Mascia24}. Several of these galaxies are extremely metal-poor (XMPs), with gas-phase metallicities $ 12 + \log\mathrm{O/H} < 8.0$ and have hard ionizing radiation \citep[][]{Jones20, Schaerer22, Rhoads23, Langeroodi23, Topping24, Saxena24, Atek24}. And importantly, they are found to be strong \Lya{} emitters (LAEs) and also produce a significant amount of ionizing radiation to easily reionize the Universe \citep{Simmonds23, Simmonds24, Saxena24, Atek24}.

Despite these significant advances in our understanding of high-$z$ galaxies, the physical mechanisms responsible for their extreme star-forming properties and high LyC escape responsible for reionization remain poorly understood due to a lack of multi-wavelength studies, which is extremely difficult with the current generation telescopes. Moreover, due to optically thick inter-galactic medium it is extremely difficult to detect LyC escape from high-$z$ galaxies \citep{Inoue14}. We thus have to rely on the local analogs of such high-$z$ galaxies with similar physical properties, e.g., low-mass, compact sizes, high sSFR, low-metallicity and strong \Lya{} emission etc. \citep[e.g.,][]{cardamone2009, yang2017, Izotov11, Izotov21a}, generally found in the Sloan Digital Sky Survey (SDSS) data. These low-$z$ ($z \sim 0.3$) analogs are known to produce a significant amount of LyC photons \citep{Schaerer16, Izotov21a} and importantly a large fraction of them leak \citep{Borthakur14, Izotov16b, Izotov16c, Izotov18a, Izotov18b-vsep, Izotov21b}. These successes in finding local LyC leakers or emitters (LCEs) have led to the recent systematic study of the Low-$z$ Lyman Continuum Survey \citep[LzLCS;][]{Flury22a} which has revealed interesting relations between the LyC escape fraction (\fesc{}) and several key galaxy physical properties \citep{Flury22b, Saldana-Lopez22, Xu23, Jaskot24-multivariate, Bait24a}. 

Recently, a rare population of extreme star-forming galaxies (\xsfg{s}) were discovered with the SDSS which are very compact XMPs having even lower stellar mass ($\log$ \mstar{} $\sim 5 - 8$), high sSFRs ($\sim 300~ $Gyr$^{-1}$) and high [OIII] $\lambda5007$ /[OII] $\lambda3727$ ratios; \Oratio $\geq 5$, with several in the range of 20-50 \citep{Izotov17b, Izotov20-diverseLyA, Izotov24-LyA-metalpoor, Jaskot19, Kouroumpatzakis24_BB}. They show diverse \Lya{} properties with several sources showing narrow \Lya{} peak velocity separation (\vsep{}) \citep{Izotov20-diverseLyA, Izotov24-LyA-metalpoor}. Such a narrow \vsep{} suggests that they are also strong LyC leakers based on the tight relation between \vsep{} and \fesc{} \citep{Verhamme15_vsep, Izotov18b-vsep}. Here we focus on the multi-frequency radio continuum (RC) observations of \xsfg{s} to gain new insights on the physical mechanisms driving their extreme SF and potentially strong \fesc{}.

RC emission from galaxies in the megahertz (MHz)- to gigahertz (GHz) frequency range has provided important insights on the star-formation process \citep{condon1992}. RC emission has been found to be an excellent tracer of extinction-free star-formation rate (SFR) in galaxies \citep[e.g.,][ and references therein]{murphy2011}. The RC emission in normal SF galaxies consists of a combination of a flat thermal emission component (free-free emission with a spectral index of $-0.1$) and a non-thermal emission component (non-thermal spectral index \alphanth{} $\sim -0.8$). The non-thermal component arises from synchrotron emission from cosmic-ray particles (CRs) accelerated by supernova (SNe) under magnetic fields. Thus the non-thermal emission can be used to infer the SNe rate, cosmic-ray (CR) energy distribution and the equipartition magnetic fields. The radio spectral energy distribution (\rcsed{}) of normal SF galaxies shows a power-law behaviour due to a combination of the thermal and non-thermal component. It is generally found that at $\sim 1.4$ GHz the non-thermal component is dominant comprising of almost 90\% of the emission component \citep[e.g.,][]{Niklas97_thermal_frac, tabatabaei2017}. Detailed multi-frequency \rcsed{} analysis of nearby SF galaxies has allowed to constrain the relative fraction of thermal (\fth{}) and non-thermal emission, and \alphanth{} \citep{tabatabaei2017} and its dependence on various galaxy physical parameters (e.g., the SFR surface density, magnetic fields etc.). Additionally in luminous and ultraluminous infrared galaxies (ULIRGs), there can be a turnover at lower frequencies in the \rcsed{} due to free-free absorption (FFA) effects \citep[e.g,][]{Clemens10, Galvin18}. Using a combination of thermal and non-thermal radio emission and FFA model, these studies find a high emission measure (EM) in ULIRGs in the range of $\sim 10^6 - 10^7$ pc cm$^{-6}$ \citep{Clemens10, Galvin18, Dey24_ULIRGS}.

RC studies of dwarf galaxies have revealed different properties compared to massive SF galaxies. The thermal fraction (\fth{}) is higher in dwarf galaxies, particularly in blue compact dwarfs \citep{Klein91, Deeg93, Thuan2004, ramya2011}. A \rcsed{} study of nearby dwarf galaxies from $\sim 0.3-24$ GHz have shown that they cannot be described by a single power-law and instead show a cuttoff/break at higher frequencies due to various physical processes related to the lack of CRs \citep{klein2018}. More relevant to our study here, for example in an extremely metal-poor compact dwarf starburst, SBS 0335-052, the integrated \rcsed{} shows a strong turnover at GHz frequencies, arising due to FFA with a EM $\sim 2 - 8 \times 10^7$ pc cm$^{-6}$ and compact size ($\sim 17$ pc) \citep[][hereafter: \citetalias{Hunt04}]{Hunt04}. Such a high EM is more commonly observed in resolved radio studies of ultra dense \hii{} regions comprising of young deeply embedded star clusters \citep[e.g.,][]{Kobulnicky99, Johnson03}. Previous single frequency (at $1.4$ GHz and $150$ MHz) RC observations of objects similar to \xsfg{s} (e.g., green-peas, blueberries) have found that these systems have lower RC flux than that expected from the standard RC-SFR relation \citep{chakraborti2012, Sebastian19, Borkar24-BB-LOFAR}. However, due to limited frequency coverage it is difficult to discern the exact physical mechanisms responsible for such a suppression. Finally, \citet{Bait24a} in a recent radio follow-up study of a sub-sample of LzLCS sources found that the \fesc{} (and the \Oratio{}, SFR density) depends on the radio spectral index (specifically measured between $3 - 6$ GHz). These results motivate a detailed \rcsed{} study of LCEs to better understand the physical processes governing these relations.

Here, we present new multi-frequency \rcsed{} observations of a sample of eight \xsfg{s} selected from \citet{Izotov20-diverseLyA, Izotov24-LyA-metalpoor} and \citet{Jaskot19}.  Our multi-frequency RC observations were conducted using the NRAO\footnote{The National Radio Astronomy Observatory is a facility of the National Science Foundation operated under cooperative agreement by Associated Universities, Inc.} Karl G. Jansky Very Large Array (\vla) at 6 (C-band), 3 (S-band), and  (a subsample) at 1.5 (L-band) GHz. We also combine new observations with the upgraded Giant Metrewave Radio Telescope (uGMRT) at Band-5 (1060 - 1460 MHz) together with archival LOw Frequency ARray (LOFAR) data at 120-168 MHz data using the LOFAR Two-metre Sky Survey Data-Release 2 \citep[LoTSS DR2;][]{Shimwell17_LoTSS-DR1, Shimwell22_LoTSS-DR2}. 

Our paper is organized as follows. In Section \ref{sec: data} we describe our sample, VLA and uGMRT observation setup, data analysis and RC flux density measurements in different radio bands. \revtext{In Section \ref{sec: results} we present the observed \rcsed{} of our sample and model the \rcsed{s} using Bayesian methods.} We then present the dust attenuation measurement using our radio observations. In Section \ref{sec: discussion}, \revtext{we discuss, 1) on the lack of non-thermal emission in \xsfg{s}} 2) the relationship between \fesc{}, \Oratio{}, metallicity, and radio spectral index for \xsfg{s} and 3) the overall implications of our observed \rcsed{} to better understand the extreme nature of high-$z$ galaxies found using the JWST. We then summarize our main results and conclude in Section \ref{sec: conclusions}.


\section{Data}
\label{sec: data}

In this section, we describe our sample selection and new \vla{} observations at multiple observing bands, namely: the S (2 - 4 GHz), C (4 - 8 GHz), X (8 - 12 GHz) and Ku (12-18 GHz) bands. We support these observations with new low frequency data using the \gmrt{} at Band-5 (1060 - 1460 MHz). We also present archival \lofar{} data (120 - 168 MHz) covering most of our sample. We also describe our calibration and imaging strategy. 

\subsection{Sample}
\begin{table*}
\caption{Basic properties of our sample of \xsfg{s}.} 
\label{table: basic properties}
\begin{center}

\begin{tabular}{ccccccccc}
\hline 
Target & RA & DEC & $z$ & $12+\log\mathrm{O/H}$ & $\log$ \mstar{} & \Oratio{}$^a$ & SFR$^b$ & Lit. source \\

     & (deg) & (deg) &  & &  & & (M$_\odot$/yr) & \\
\hline \hline  \\
J0159+0751 & $29.96979$ & $+7.86356$ & $0.06105$ & $7.56$ & $8.6$ & $39.0$ & $2.3$ & \citet{Izotov20-diverseLyA} \\
J0811+4730 & $122.96721$ & $+47.50729$ & $0.0444$ & $6.98$ & $5.88$ & $5.9$ & $0.19$ & \citet{Izotov24-LyA-metalpoor} \\
J0820+5431 & $125.08029$ & $+54.52781$ & $0.03851$ & $7.49$ & $6.0$ & $22.0$ & $0.2$ & \citet{Izotov20-diverseLyA}\\
J1032+4919 & $158.23633$ & $+49.32979$ & $0.0442$ & $7.59$ & $6.8$ & $24.0$ & $2.4$ & \citet{Izotov20-diverseLyA} \\
J1355+4651 & $208.85692$ & $+46.86426$ & $0.02811$ & $7.56$ & $5.8$ & $23.0$ & $0.1$ & \citet{Izotov20-diverseLyA} \\
J150934+373146 & $227.39239$ & $+37.52948$ & $0.0325$ & $7.88$ & $8.1$ & $15.1$ & $1.35$ & \citet{Jaskot19} \\
J160810+352809 & $242.04318$ & $+35.46926$ & $0.0327$ & $7.83$ & $7.5$ & $34.9$ & $0.46$ & \citet{Jaskot19} \\
J2229+2725 & $337.38829$ & $+27.42378$ & $0.07622$ & $7.08$ & $6.96$ & $53.0$ & $0.68$ & \citet{Izotov24-LyA-metalpoor} \\
\hline
\end{tabular}

\hbox{Notes. The physical properties presented here are taken from the literature values shown in the last column. }
\hbox{$^a$ \Oratio{} $\equiv$ [OIII] $\lambda5007$ /[OII] $\lambda3727$. }
\hbox{$^b$ The SFR is derived using the dust-corrected \hbeta{} line using the relation from \citet{Kennicutt98}. For \bbten{} and \bbthirteen{} the H$\alpha$ line was} \hbox{used and the relation from \citet{Kennicutt12}. See the references (last column) for details.}
\end{center}
\end{table*}

\begin{table*}
    \centering
    \caption{\revtext{Summary of the new radio observations of \xsfg{s}.}}
    \label{table: obs_log}
    \begin{tabular}{lcccccc}
        \hline
        \hline
        \multirow{2}{*}{Source ID} & \multicolumn{2}{c}{VLA Observations} & \multirow{2}{*}{VLA Project ID} & \multirow{2}{*}{GMRT Project ID$^{a}$} & \multirow{2}{*}{LOFAR} \\
        \cline{2-3}
        & B-Config. & C-Config. & & & \\
        \hline
        J0159+0751 & S, C & X & 23A-138, 24A-103 & 43\_062 & $--$ \\
        \hline
        J0811+4730 & S, C & X, Ku & 23A-138, 24A-103 & 43\_062 & LoTSS DR2 \\
        \hline
        J0820+5431 & S, C & X & 23A-138, 24A-103 & 45\_075 &  LoTSS DR2 \\
        \hline
        J1032+4919 & L, S, C & X, Ku & 23A-138, 24A-103 & 43\_062 &  LoTSS DR2 \\
        \hline
        J1355+4651 & S, C & X, Ku & 23A-138, 24A-103 & 45\_075 &  LoTSS DR2 \\
        \hline
        \bbten{} & S, C & X, Ku & 23A-138, 24A-103 & 34\_123 (from \citetalias{Sebastian19}) & LoTSS DR2 \\
        \hline
        \bbthirteen{} & S, C & X, Ku & 23A-138, 24A-103 & 34\_123 (from \citetalias{Sebastian19}) & LoTSS DR2 \\
        \hline
        J2229+2725 & S, C & X, Ku & 23A-138, 24A-103 & 45\_075 &  LoTSS DR2 \\
        \hline
    \end{tabular}
    
    \hbox{\revtext{Notes. $^{a}$ All the \gmrt{} observations used here are performed in Band-5.}}
        
\end{table*}

Our sample is selected from the recently identified sample of \xsfg{s} at low-$z$, ~$z \sim 0.01 - 0.06$, from \citet{Izotov20-diverseLyA, Izotov24-LyA-metalpoor} and \citet{Jaskot19} \citep[see also][]{yang2017}. All our sources have extremely low-metallicity ($12+\log\mathrm{O/H} \sim  6.98 - 7.88$), low stellar mass ($\log$ (\mstar{}) $\sim 5.8 - 8.6$), very high \Oratio{} ($\gtrsim 15$ to up to $\sim 53$) and high \hbeta{} equivalent width (\ewhbeta{} $\gtrsim 250~\AA$). \revtext{Table \ref{table: basic properties} summarizes the basic physical properties of our sample of \xsfg{s}.} We observed several of these sources at L-, S- and C-bands using the \vla{} with the B-configuration (PI: Bait; Project Code: 23A-138 \& 24A-103) and at higher frequency bands at X- and Ku with the C-configuration (PI: Bait; Project Code: 24A-103). We also conducted follow-up \gmrt{} Band-5 observations (PI: Bait, Proposal ID: 43\_062 and 45\_075) for our sample of \xsfg{s}. \revtext{Table \ref{table: obs_log} summarizes the observations in different bands and the corresponding project IDs.}

\revtext{Using these data allows us to study the \rcsed{} at GHz frequencies for these sources.} We thus present a sample of 8 \xsfg{s} which has a seamless RC coverage between $\sim 1 - 12$~GHz (S- to X- band) from the two \vla{} and \gmrt{} programs. Except for one source (J0159+0751), all our sources also have Ku-band ($12 - 18$~GHz) observations (Project Code: 24A-103). For two sources, \bbten{} and \bbthirteen{}, we use the published Band-5 data from \citet{Sebastian19} (\citetalias{Sebastian19} hereafter). J1032+4919 was also observed with the \vla{} at L-band ($1 - 2$~GHz; under Project Code: 23A-138). 
We also present \lofar{} 150 MHz data which covers 7 out of the 8 sources from our sample. The various physical properties (e.g., metallicity, \Oratio{}, \vsep{}, observed and dust extinction corrected \hbeta{} line flux) of the 8 \xsfg{s} studied here are taken from \citet{Izotov20-diverseLyA, Izotov24-LyA-metalpoor} and \citet{Jaskot19} and shown in Table \ref{table: basic properties}.

In the next section, we present details on the new RC observations at multiple \vla{} and \gmrt{} bands together with archival \lofar{} observations. Table \ref{table: obs summary} provides a brief summary of these observations.

\subsection{VLA L-, S-, C-, X- and Ku-band Observations and Data Analysis}
\label{sec: vla data}
We used the default \vla{} setup for our L-, S-, and C-band continuum observations. For the L- and S-bands we used a total bandwidth of 1 and 2 GHz respectively with a 8-bit sampler. The L-band (S-band) was split into one (two) baseband(s) each comprising of 8 $\times$ 128 MHz subbands. For the C-band we used a total bandwidth of 4 GHz with a 3-bit sampler which was split into two basebands each comprising of 16 $\times$ 128 MHz subbands. For each of these cases, the subbands were further split into 64 channels. All of these observations were performed in the B-configuration. This setup is identical to the previously followed setup for RC observations of the LzLCS sources at L-, S- and C-bands from \citet{Bait24a}. We adopted the standard observing strategy used for continuum observations at L-, S- and C-bands which was also followed for the LzLCS sources \citep[see][for details]{Bait24a}. Briefly, after an initial setup, we observed the flux density scale calibrator (for $3 -5$ mins). We then alternative between the phase calibrator ($\sim 1 $min per cycle) and the target ($\sim 5 - 6$ mins per cycle). The S- and C-band observations were conducted within a single scheduling block (SB) lasting approximately $\sim$1~hr. We achieved an on-source integration time of 22 mins each for S- and C-band. For the L-band SBs typically lasted for a duration of $\sim$1~hr and 40 mins, where we achieved an on-source integration time of $\sim 75$~mins. 

For the high-frequency X- and Ku-band observations we again used the default \vla{} setup, with a total bandwidth of 4 GHz and 6 GHz respectively with a 3-bit sampler. The total bandwidth for the X-band (Ku-band) was split in 2 (3) subbands each comprising 16 $\times$ 128 MHz subbands. Here as well each subband was further split into 64 channels. These observations were performed in the C-configuration. After an initial setup, we observed the flux density scale calibrator for approximately 5 mins. For the Ku-band an additional scan of $\sim$4 mins for pointing calibration was performed, where the phase calibrator was used as a pointing calibrator. We then alternative between the phase calibrator ($\sim 2$ mins per cycle) and the target ($\sim 8$ mins per cycle). We achieved an on-source integration time of $\sim 15$ mins for both the X- and Ku-band. 

For all our \vla{} data reduction, we used the Common Astronomy Software Applications (\textsc{CASA}) data processing software \citep{mcmullin2007casa, CASA}. We followed a similar data reduction procedure as described in \citet{Bait24a}. Briefly, each SB was initially flagged and calibrated using the standard VLA calibration pipeline v6.5.4. We manually inspected the calibrated flux density scale and phase calibrators for each SB. Any leftover RFI and poorly performing antennas were manually flagged. And the \vla{} pipeline was then re-run. 

We then extracted the calibrated target and inspected for any leftover RFI in some spectral windows, which was then further flagged. Then we proceeded with the imaging of these data using CASA \textsc{tclean}. For S- and C-bands, since these were observed in the multi-band mode, we first split the target in S- and C-band before imaging \citep[see][]{Bait24a}). For all our imaging, CASA \textsc{tclean} was setup to use the multi-term multi-frequency synthesis (mtmfs) algorithm \citep{Rau11} where the \textsc{nterms} were set to 2 and a \textsc{robust} value of 0.5 was used. The flux density and corresponding error ($\sigma$) was then estimated by manually making elliptical regions around each of the target sources using CASA task \textsc{imview}, and a 2D Gaussian fit was performed using the task \textsc{imfit} \citep[see][for more details on this procedure]{Bait24a}. For one marginally resolved source in our sample, \bbten{} which is also relatively extended in the optical compared to other \xsfg{s}, the flux density and error was measured using PyBDSF \citep[the Python Blob Detector and Source Finder;][]{Mohan15}. \revtext{Table \ref{table: radio fluxes} presents the flux density and noise ($\sigma$) in each band for the sources in our sample. For non-detections, where sources have a signal-to-noise ratio (S/N) below 3, we present the 3$\sigma$ upper limits as the flux density.} Table \ref{table: radio fluxes} also shows the synthesized beam properties (beam major/minor axis and position-angle) from our observations in various bands. Note that for all further analysis on the RC flux density, spectral index and \rcsed{} analysis we add a 5\% calibration error in quadrature \citep[following][]{Bait24a}. 

\subsection{\gmrt{} Band-5 Observations and Data Analysis}
\label{sec: gmrt data}
For the \gmrt{} Band-5 (1460 - 1060 MHz) observations, we used a total bandwidth of 400 MHz split in 2048 channels, similar to that used in \citetalias{Sebastian19}. 

The data reduction was carried out using the \yarp{}\footnote{The current development version of the \yarp{} package can be found here: \url{https://github.com/omkarbait/yarp/}} package. \yarp{} comprises a collection of high-level \python{} recipes to perform flagging, calibration and imaging which are based on various standard CASA tasks at the lowest-level. The different flagging, calibration and imaging recipes can be combined to design custom radio data reduction pipelines/workflows for both continuum and spectral line data. The \yarp{} package was first used for GMRT \hi{} 21-cm spectral-line data reduction in \citet{Paudel23}, and a detailed paper describing the package is in preparation. 

Here we use the \yarp{} package with CASA v6.5.5 to design a custom pipeline for our \gmrt{} Band-5 continuum data reduction. Briefly, in this pipeline the raw \gmrt{} data initially underwent a round of flagging using the CASA task \textsc{tfcrop}. Here any bad/non-working antennas were also flagged which was manually provided to \yarp{}. This data then underwent 5 rounds of calibration (flux, phase and bandpass) and flagging (which used a combination of CASA tasks \textsc{tfcrop, rflag} and \textsc{clip}). The calibrated data for phase calibrator, flux density scale calibrator and target was manually inspected for any leftover RFI. We then extracted the calibrated target and self-calibrated the data. Each self-calibration round includes three rounds of phase-only self-calibration and one round of Amplitude and Phase self-calibration.  For a few cases, the residual (data - model) was flagged using CASA task \textsc{rflag} and the flags were applied to the data before proceeding to another round of self-calibration. This final calibrated data was then used for imaging using the {\sc mtmfs} algorithm with \textsc{nterms} set to 2. For all our new Band-5 observations presented here, we do not detect any emission and thus we present the $3\sigma$ upper limits. For two \xsfg{s}, \bbten{} and \bbthirteen{}, the flux density and noise are taken from \citetalias{Sebastian19}. Note that J1032+4919 was also observed by \citetalias{Sebastian19} and was also not detected. Our upper limits for J1032+4919 are consistent with \citetalias{Sebastian19}. Table \ref{table: radio fluxes} presents the flux properties for our sample.

\subsection{Archival LoTSS DR2 data}
\label{sec: lofar data}
We also present archival LoTSS (120-168 MHz) DR2 imaging data \citep{Shimwell22_LoTSS-DR2} for our sample. We download cutouts around each of the sources in our sample using the LoTSS DR2 cutout service \footnote{\url{https://lofar-surveys.org/dr2_release.html}}. We inspected each image to search for detections. Out of the 7 sources which have LoTSS coverage, we detected emission in only one source, \bbten{}. The flux was estimated using PyBDSF. For the rest of the sources we present the  $3\sigma$ upper limits (see Table \ref{table: radio fluxes}).

\begin{table}
\caption{Brief summary of the observation setup for the new multi-frequency radio observations.} 
\label{table: obs summary}
\begin{center}
\begin{tabular}{cccccc}
\hline
Tel. & Band    & Cen. Freq.$^a$ & BW$^b$   & TOS$^c$ & Array$^d$ \\
     &         & (GHz)      & (GHz) & (mins)           &       \\
     \hline \hline \\
\gmrt{} & Band-5  & 1.25       & 0.4   & 120            & Full  \\
\vla{}  & L   & 1.5        & 1     & 75             & B     \\
\vla{}  & S$^e$  & 3.0        & 2     & 20             & B     \\
\vla{}  & C$^e$  & 6.0        & 4     & 20             & B     \\
\vla{}  & X  & 10.0       & 4     & 15             & C     \\
\vla{}  & Ku & 15.0       & 6     & 15             & C    \\
\hline 
\end{tabular}

\hbox{Notes.\ $^a$ Central frequency of the observing band. }
\hbox{$^b$ Total Bandwidth. } 
\hbox{$^c$ Time on source (approximate values). }
\hbox{$^d$ \vla{} configuration.}
\hbox{$^e$The S- and C-band observations were performed simultaneously in a multi-band mode. }
\end{center}
\end{table}


\begin{landscape}
\begin{table}
\begin{center}
\caption{Radio Fluxes of \xsfg{}.} 
\label{table: radio fluxes}

\begin{tabular}{cccccccccc}
\hline
 Freq. (GHz)  & Details$^{a}$ & J0159+0751 & J0811+4730 & J0820+5431 & J1032+4919 & J1355+4651 & J150934+373146 & J160810+352809 & J2229+2725 \\
\hline \hline \\
  & $S_\nu$ ($\mu$Jy) & $< 64.5$ & $< 58.2$ & $< 69.3$ & $< 48.3$ & $< 46.2$ & $478\pm36.9^b$ & $58\pm17.0^b$ & $< 79.8$ \\
1.25 & $\theta_{\mathrm{maj}} \times \theta_{\mathrm{min}}$ & $2.3\prime\prime \times 2.1\prime\prime $ & $2.4\prime\prime \times 1.9\prime\prime $ & $5.0\prime\prime \times 2.3\prime\prime $ & $2.4\prime\prime \times 2.2\prime\prime $ & $4.0\prime\prime \times 2.1\prime\prime $ & $3.1\prime\prime \times 2.6\prime\prime $ & $7.3\prime\prime \times 3.3\prime\prime $ & $3.9\prime\prime \times 2.0\prime\prime $ \\
  & P.A.  & $56.2^\circ$ & $21.1^\circ$ & $85.4^\circ$ & $7.4^\circ$ & $-35.6^\circ$ & $-69.7^\circ$ & $55.4^\circ$ & $44.0^\circ$ \\
\hline \\ 

  & $S_\nu$ ($\mu$Jy) & $--$ & $--$ & $--$ & $< 41.1$ & $--$ & $--$ & $--$ & $--$ \\
1.5 & $\theta_{\mathrm{maj}} \times \theta_{\mathrm{min}}$ & $--$ & $--$ & $--$ & $3.8\prime\prime \times 3.5\prime\prime $ & $--$ & $--$ & $--$ & $--$ \\
& P.A. & $--$ & $--$ & $--$ & $-40.8^\circ$ & $--$ & $--$ & $--$ & $--$ \\

\hline \\ 
  & $S_\nu$ ($\mu$Jy) & $< 36.6$ & $< 27.3$ & $< 31.8$ & $68.7\pm9.6$ & $< 32.7$ & $330.0\pm39.6$ & $--^{c}$ & $< 29.7$ \\
3.0  & $\theta_{\mathrm{maj}} \times \theta_{\mathrm{min}}$ & $3.3\prime\prime \times 1.8\prime\prime $ & $2.7\prime\prime \times 1.9\prime\prime $ & $2.0\prime\prime \times 1.7\prime\prime $ & $2.3\prime\prime \times 1.8\prime\prime $ & $1.9\prime\prime \times 1.7\prime\prime $ & $6.5\prime\prime \times 1.7\prime\prime $ & $2.4\prime\prime \times 1.8\prime\prime $ & $2.3\prime\prime \times 1.8\prime\prime $ \\
  & P.A. & $55.4^\circ$ & $68.5^\circ$ & $38.6^\circ$ & $-83.8^\circ$ & $49.7^\circ$ & $59.9^\circ$ & $84.3^\circ$ & $71.1^\circ$ \\
\hline \\ 

  & $S_\nu$ ($\mu$Jy)& $34.6\pm6.4$ & $< 16.8$ & $< 18.9$ & $61.9\pm5.7$ & $< 19.8$ & $170.0\pm15.14$ & $46.45\pm4.6$ & $< 18.6$ \\
6.0  & $\theta_{\mathrm{maj}} \times \theta_{\mathrm{min}}$ & $1.6\prime\prime \times 1.0\prime\prime $ & $1.5\prime\prime \times 1.0\prime\prime $ & $1.2\prime\prime \times 0.9\prime\prime $ & $1.2\prime\prime \times 1.0\prime\prime $ & $1.1\prime\prime \times 0.9\prime\prime $ & $2.4\prime\prime \times 1.0\prime\prime $ & $1.1\prime\prime \times 1.0\prime\prime $ & $1.1\prime\prime \times 1.1\prime\prime $ \\
  & P.A. & $50.0^\circ$ & $66.0^\circ$ & $42.4^\circ$ & $-62.0^\circ$ & $43.5^\circ$ & $63.6^\circ$ & $89.7^\circ$ & $57.6^\circ$ \\

\hline \\ 

  & $S_\nu$ ($\mu$Jy)& $36.0\pm6.6$ & $< 18.4$ & $< 39.9$ & $53.9\pm6.8$ & $31.0\pm6.6$ & $210.0\pm12.16$ & $43.5\pm7.6$ & $< 16.7$ \\
10.0  & $\theta_{\mathrm{maj}} \times \theta_{\mathrm{min}}$ & $4.4\prime\prime \times 1.9\prime\prime $ & $3.4\prime\prime \times 1.9\prime\prime $ & $2.8\prime\prime \times 1.9\prime\prime $ & $3.9\prime\prime \times 1.8\prime\prime $ & $5.5\prime\prime \times 1.8\prime\prime $ & $2.0\prime\prime \times 1.9\prime\prime $ & $2.1\prime\prime \times 1.8\prime\prime $ & $2.7\prime\prime \times 1.9\prime\prime $ \\
  & P.A. & $-56.0^\circ$ & $79.4^\circ$ & $-83.4^\circ$ & $-74.6^\circ$ & $64.3^\circ$ & $41.7^\circ$ & $-19.3^\circ$ & $-75.8^\circ$ \\

\hline \\

 & $S_\nu$ ($\mu$Jy)& $--$ & $< 17.2$ & $--$ & $69.0\pm5.0$ & $28.3\pm5.4$ & $180.0\pm22.42$ & $45.14\pm6.0$ & $< 13.9$ \\
15.0 & $\theta_{\mathrm{maj}} \times \theta_{\mathrm{min}}$ & $--$ & $1.6\prime\prime \times 1.3\prime\prime $ & $--$ & $1.4\prime\prime \times 1.2\prime\prime $ & $2.0\prime\prime \times 1.2\prime\prime $ & $3.1\prime\prime \times 1.3\prime\prime $ & $2.5\prime\prime \times 1.3\prime\prime $ & $2.9\prime\prime \times 1.3\prime\prime $ \\
& P.A. & $--$ & $-70.7^\circ$ & $--$ & $7.4^\circ$ & $89.7^\circ$ & $73.1^\circ$ & $71.6^\circ$ & $67.0^\circ$ \\

\hline \\

 & $S_\nu$ ($\mu$Jy)& $--$ & $< 532.4$ & $< 169.5$ & $< 200.2$ & $< 181.9$ & $400.68\pm126.6$ & $< 225.8$ & $< 316.0$ \\
0.150 & $\theta_{\mathrm{maj}} \times \theta_{\mathrm{min}}$ & $--$ & $6\prime\prime \times 6\prime\prime $ & $6\prime\prime \times 6\prime\prime $ & $6\prime\prime \times 6\prime\prime $ & $6\prime\prime \times 6\prime\prime $ & $6\prime\prime \times 6\prime\prime $ & $6\prime\prime \times 6\prime\prime $ & $6\prime\prime \times 6\prime\prime $ \\
  & P.A. & $--$ & $90^\circ$ & $90^\circ$ & $90^\circ$ & $90^\circ$ & $90^\circ$ & $90^\circ$ & $90^\circ$ \\
\hline
\end{tabular}

\hbox{Notes.\ Sources with no observations in a specific band are shown with empty values. }
\hbox{$^a$ Shows the measured flux density ($S_\nu$) and beam parameters for each observing frequency. For non-detections we present the $3\sigma$ upper limits on the flux densities.}
\hbox{$^b$ Flux densities and beam parameters taken from \citetalias{Sebastian19}. }
\hbox{$^c$ The VLA S-band flux density for this source could not be measured due to strong artefacts from a neighbouring bright source.}
\end{center}
\end{table}
\end{landscape}



\section{Results}
\label{sec: results}

\revtext{ Here we present the observed \rcsed{s} of 5 out of 8 \xsfg{s} which show detections in one or more radio bands. These sources span a wide range of radio frequencies from 1.2 GHz (from 150 MHz for \bbten{}) to up to 15 GHz using a combination of \vla{}, \gmrt{}, and \lofar{} telescopes described in Section \ref{sec: data}. We perform a Bayesian SED modelling of these sources to derive various physical parameters. We exclude the three other \xsfg{s} from such an analysis since they do not show detections in any bands and no useful constraints can be derived on the physical parameters from their data. 
}

\begin{figure*}
\captionsetup[subfigure]{labelformat=empty}
\centering
\subfloat[]{{\includegraphics[width=0.44\hsize]{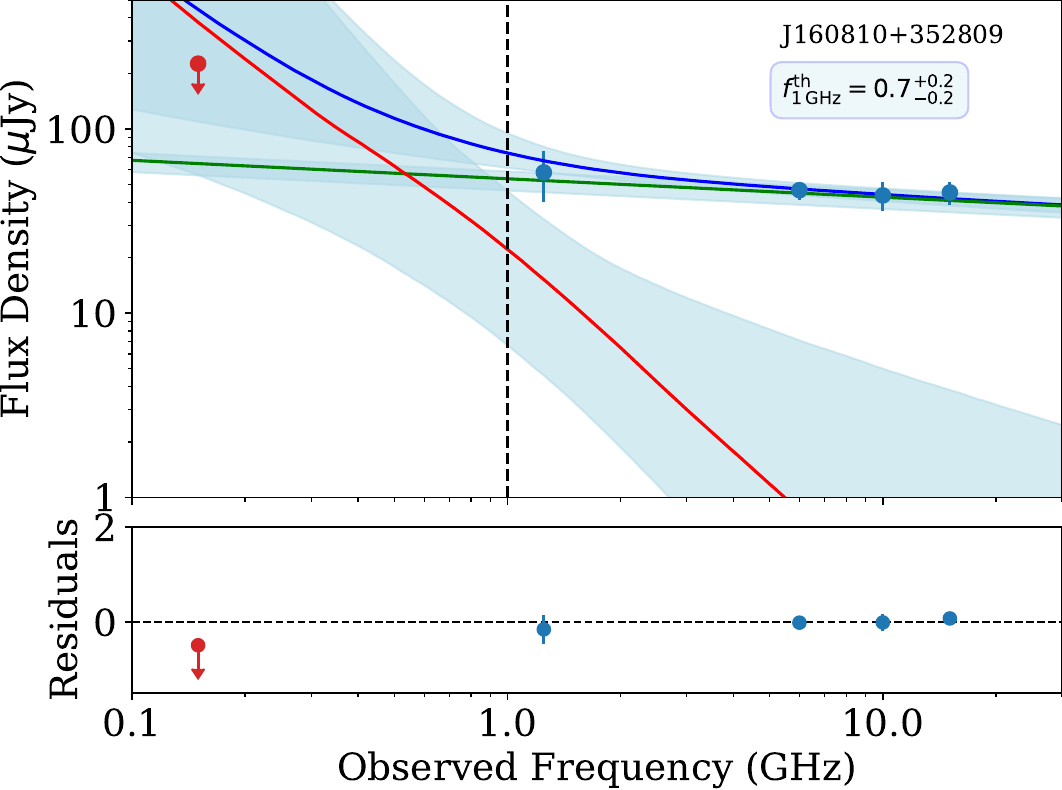 } }}%
\qquad
\subfloat[]{{\includegraphics[width=0.44\hsize]{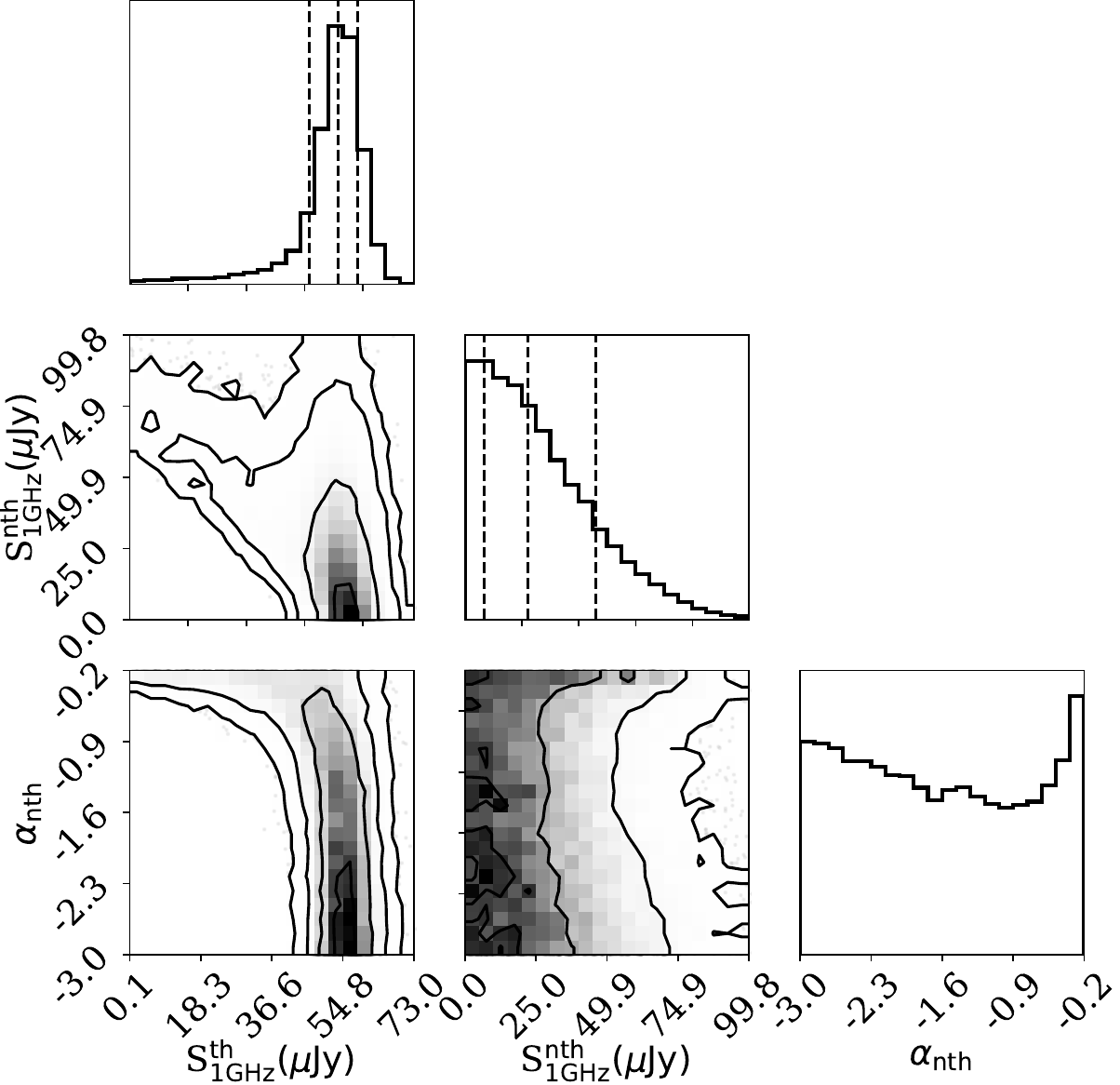 }}}%
  \qquad
\subfloat[]{{\includegraphics[width=0.44\hsize]{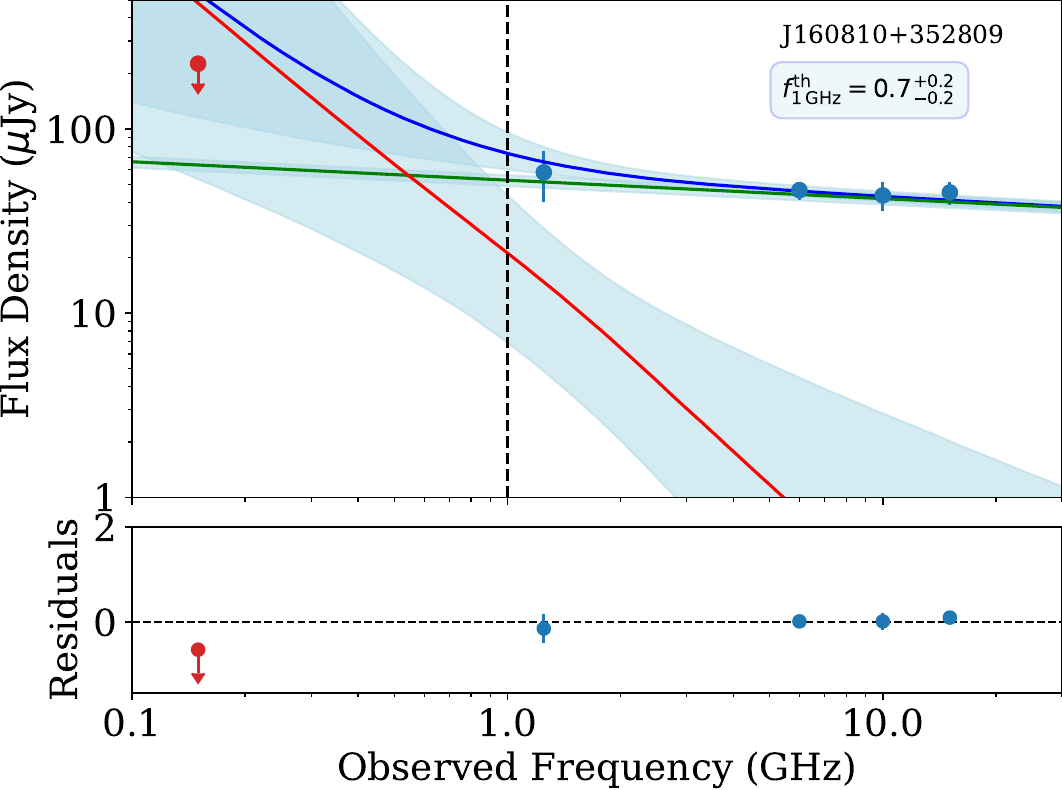} }}%
\qquad
  \subfloat[]{{\includegraphics[width=0.44\hsize]{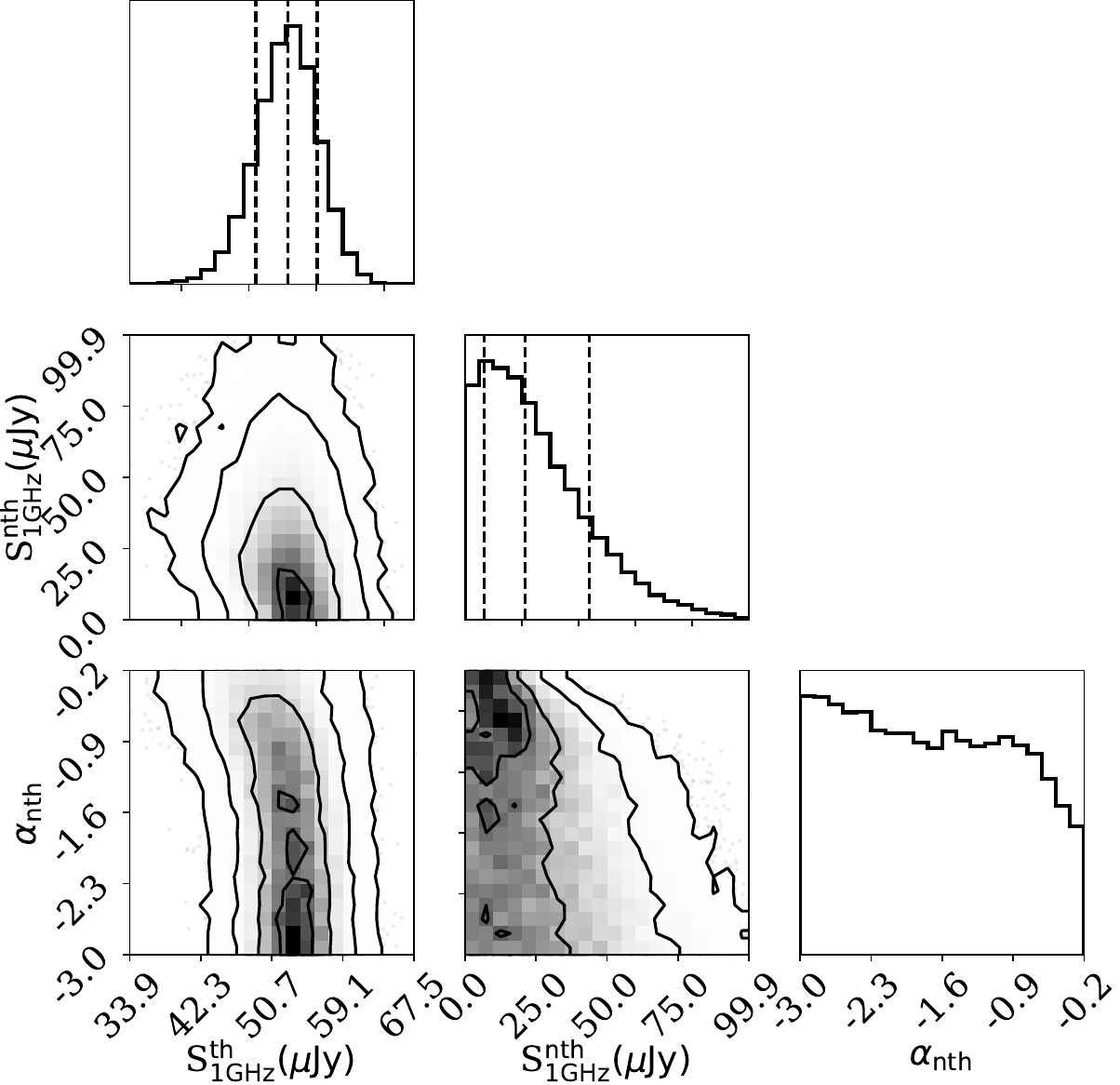 }}}%
  
  \caption{\revtext{ Bayesian \rcsed{} fitting results for \bbthirteen{} using the \revtexts{SFG-thin} model. Top (bottom) row shows the fitting with flat (Gaussian) priors on \sth{}. The left panels shows RC detections with blue error bars (wth $1\sigma$ errors), and non-detections ($3\sigma$ upper limits) are shown in red arrows. The corresponding best-fit (median) \revtexts{SFG-thin} model is shown in solid blue line. The thermal (non-thermal) components are shown in green (red) solid line. The shaded area shows the 16th and 84th percentile range derived from the MCMC samples. We show the fraction residuals ( (data - best-fit model)/best-fit model) in the lower subplot in the left column. The right column shows the 2D and 1D marginalized posteriors on \sth{}, \snth{} and \alphanth{} derived from the MCMC chains. The contours enclose $19.3\%, 68\%, 95\%$ and $99.7\%$ of the 2D marginalized posterior probability. \revtexts{The dashed lines on the 1D marginalized posteriors show the $16$th, $50$th (median) and $84$th percentiles. } See Section \ref{sec: radio-sed individual cases} and Table \ref{table: bayes fitting} for details on the priors, choice of model and the bands used for the Bayesian fitting. Table \ref{table: bayes parameters} summarizes the constraints on the various parameters and constraints on \fth{}. } }

  \label{fig: BB13 bayes rado-sed fit}
\end{figure*}


\begin{figure*}
\captionsetup[subfigure]{labelformat=empty}
\centering
\subfloat[]{{\includegraphics[width=0.44\hsize]{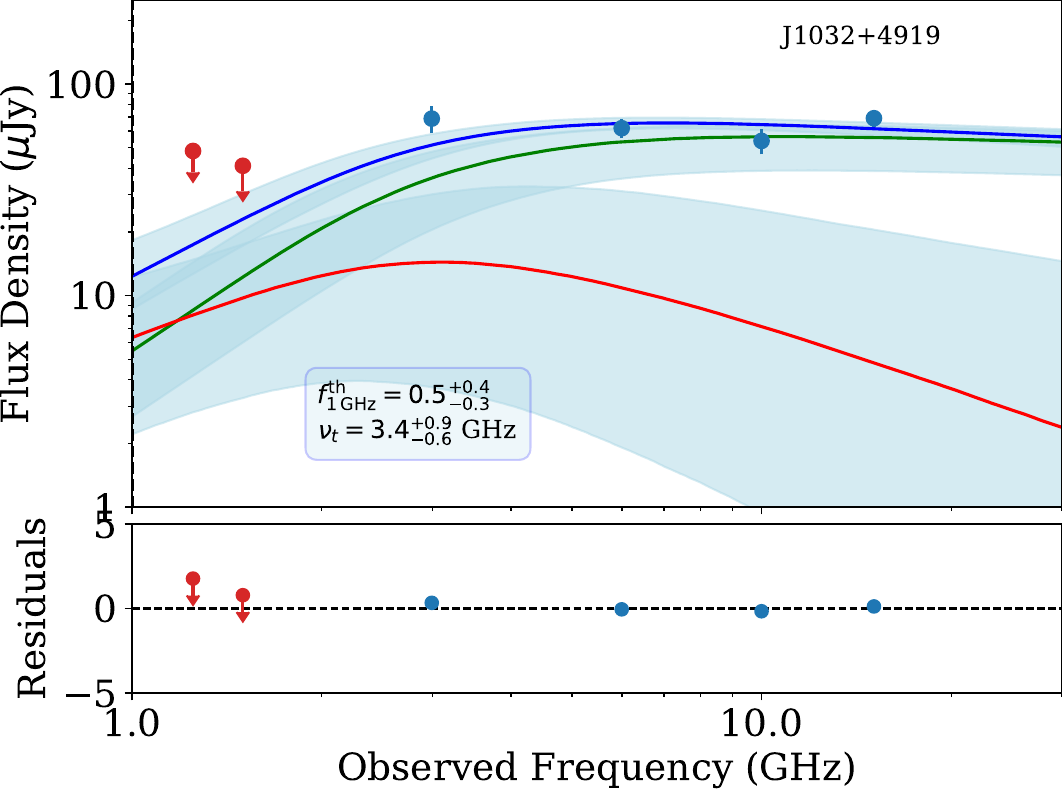} }}%
\qquad
\subfloat[]{{\includegraphics[width=0.44\hsize]{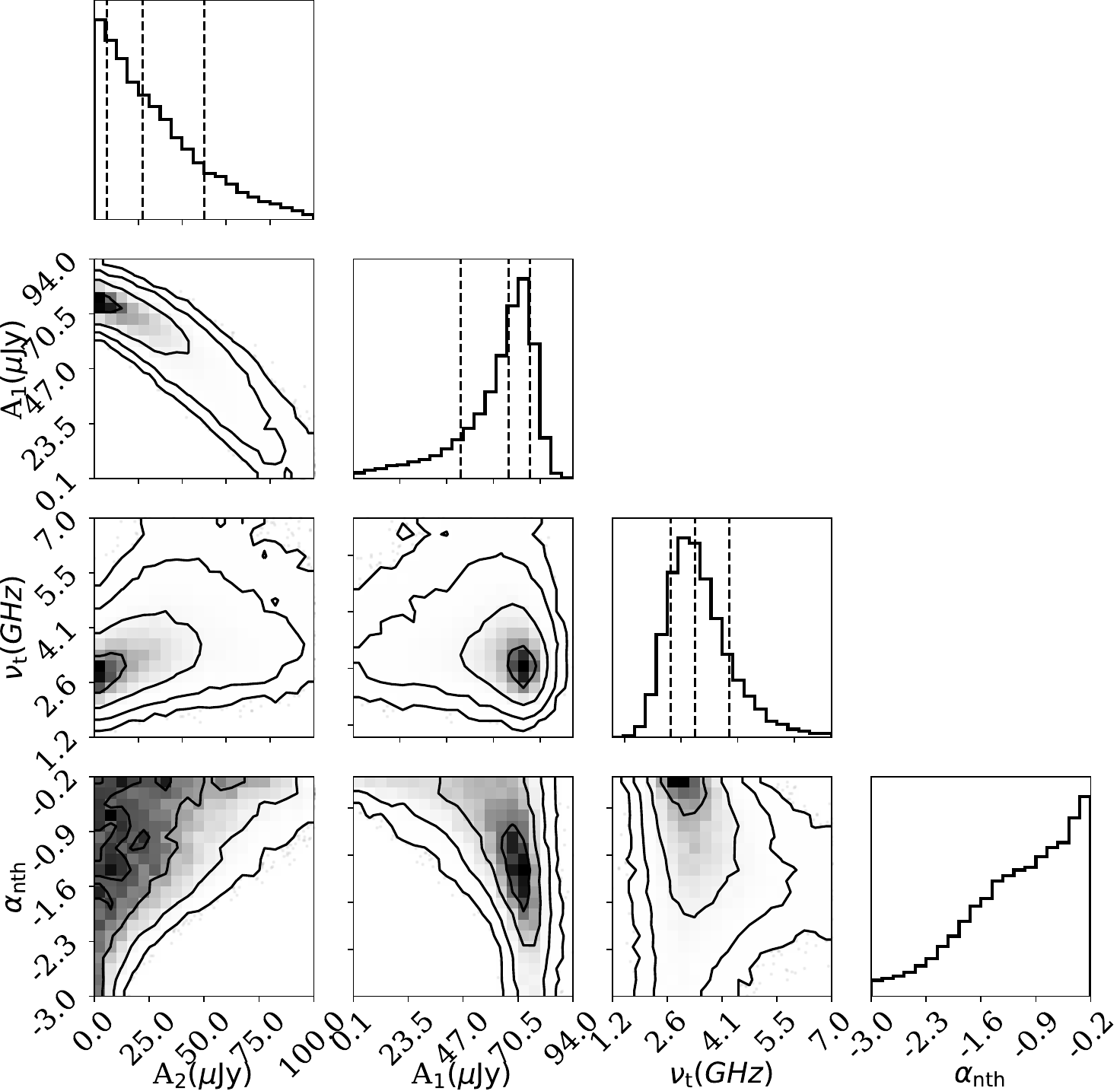 }}}%
  \qquad
\subfloat[]{{\includegraphics[width=0.44\hsize]{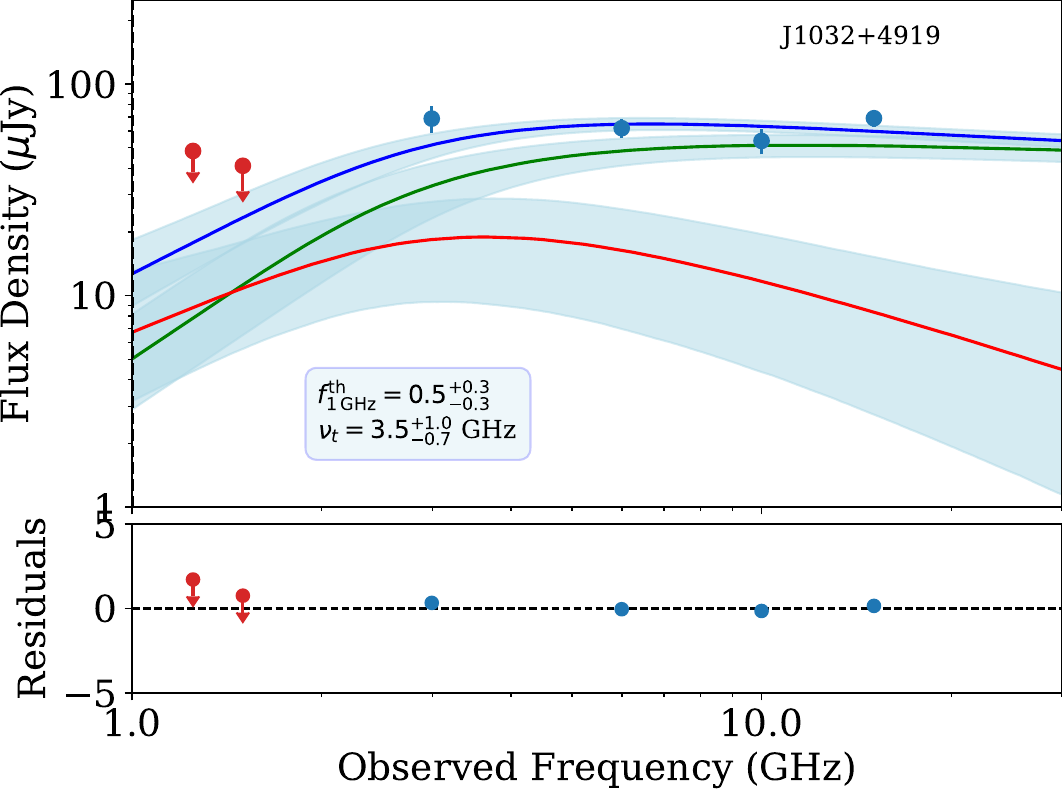} }}%
\qquad
  \subfloat[]{{\includegraphics[width=0.44\hsize]{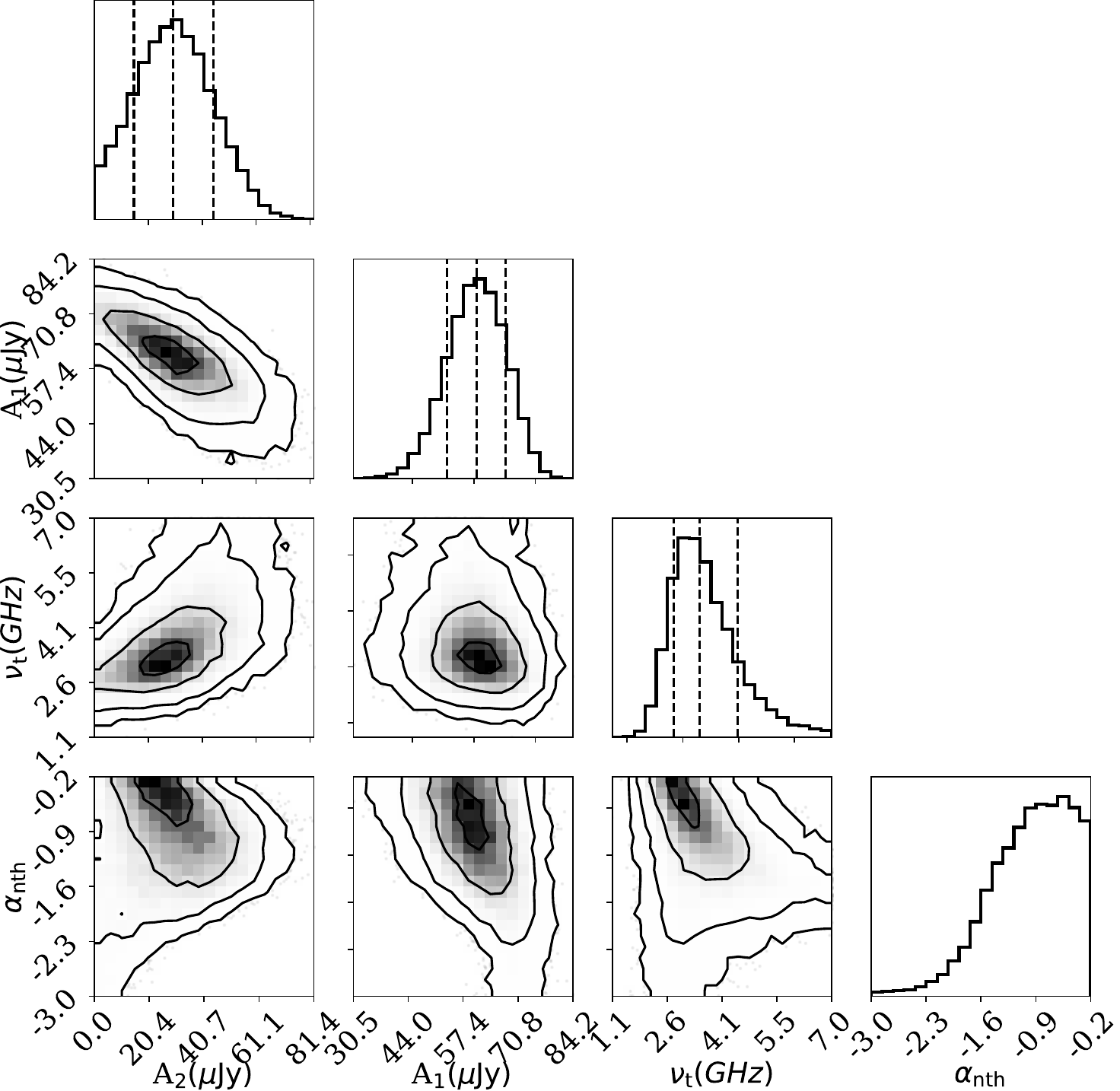 }}}%
  \caption{\revtext{ Bayesian \rcsed{} fitting results for J1032+4919 using the \revtexts{SFG-thick} model. The various panels follow the same order as Figure \ref{fig: BB13 bayes rado-sed fit} in terms of the choice of priors on \sth{}. We constrain the \nuturn{} to $\sim 3$GHz across the different choice of priors (Table \ref{table: bayes parameters}). Overall we find that the \rcsed{} is well fit with a thermally dominant \revtexts{SFG-thick} \rcsed{} with a turnover at $\sim 3$ GHz. See Section \ref{sec: radio-sed individual cases} for a detailed discussion on this source.} }

  \label{fig: J1032 bayes rado-sed fit}
\end{figure*}

\begin{figure*}
\captionsetup[subfigure]{labelformat=empty}
\centering
\subfloat[]{{\includegraphics[width=0.44\hsize]{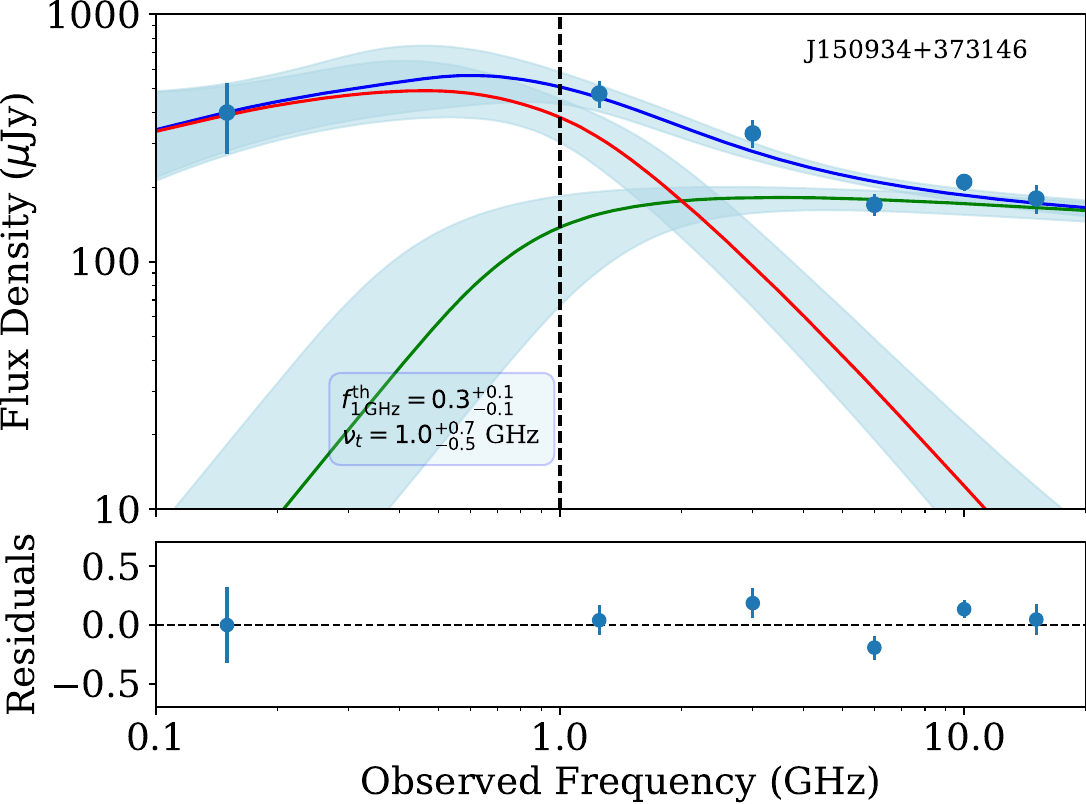} }}%
\qquad
\subfloat[]{{\includegraphics[width=0.44\hsize]{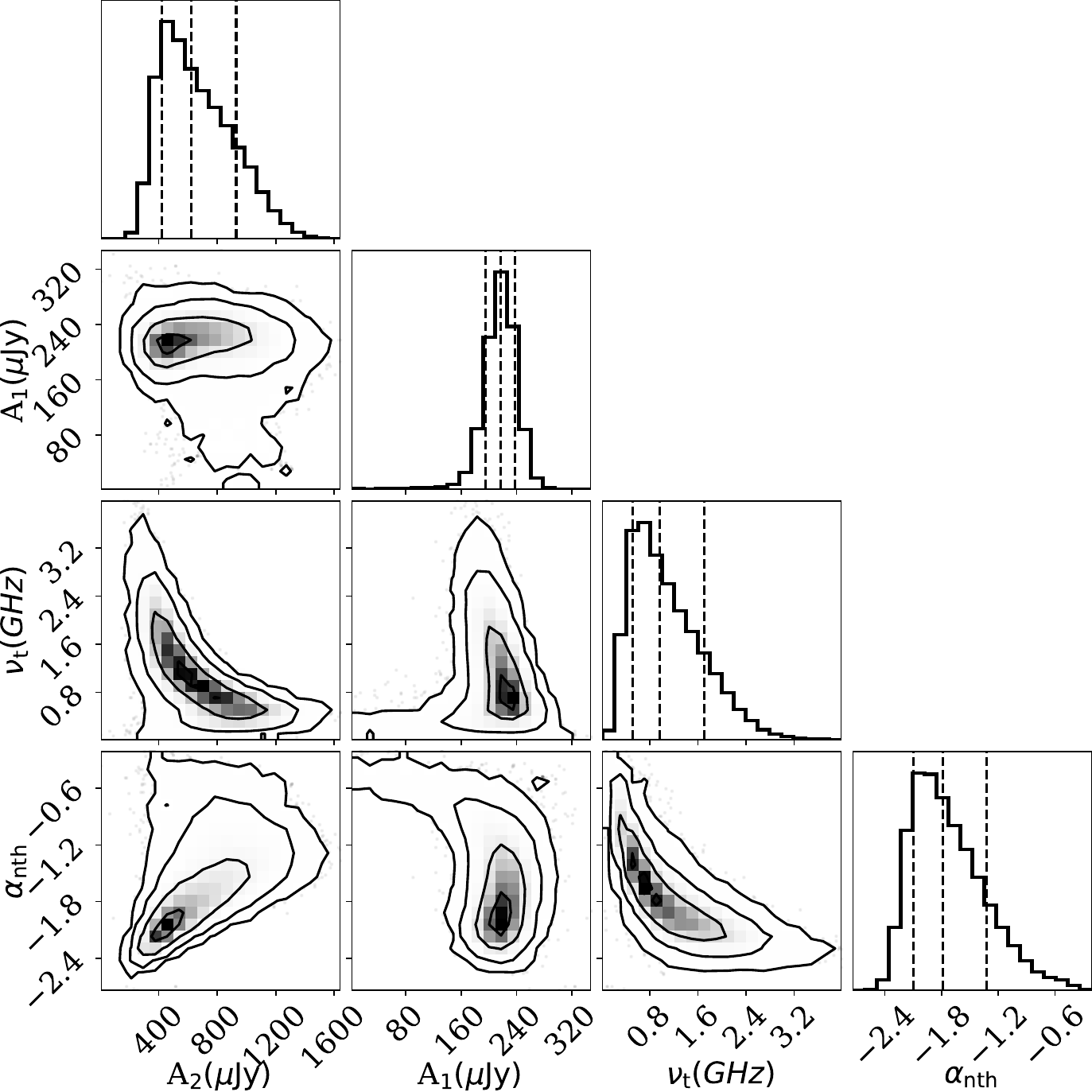 }}}%
  \qquad
\subfloat[]{{\includegraphics[width=0.44\hsize]{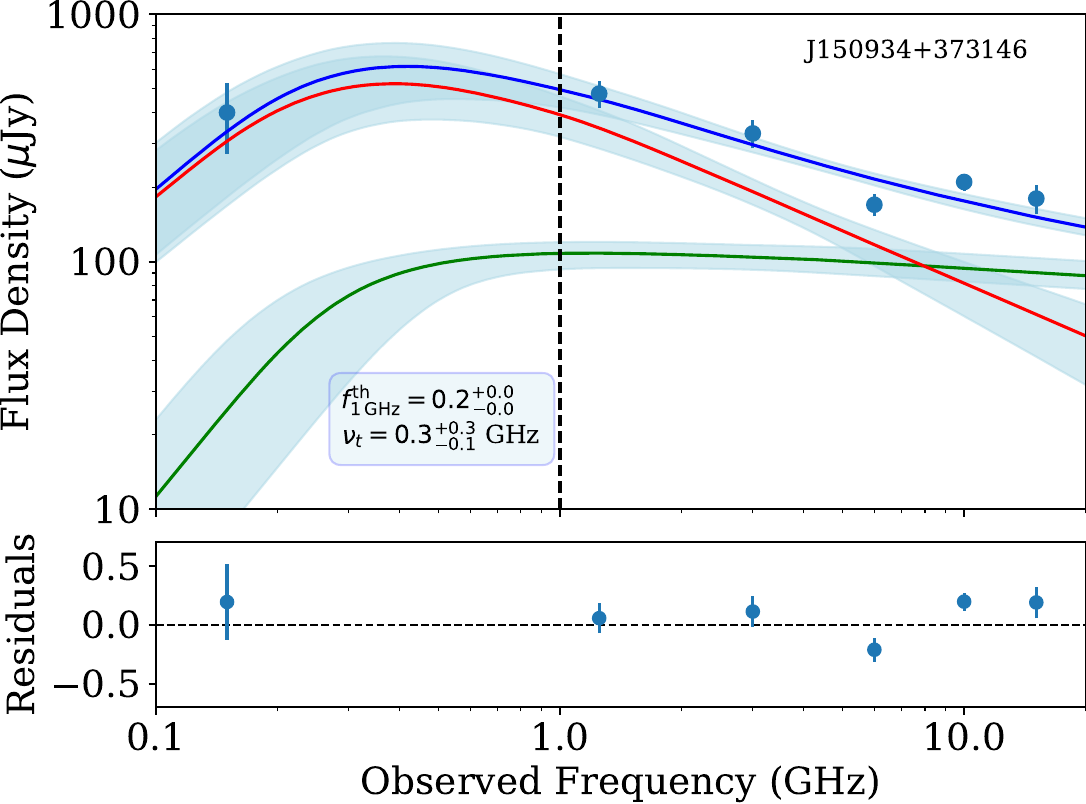} }}%
\qquad
  \subfloat[]{{\includegraphics[width=0.44\hsize]{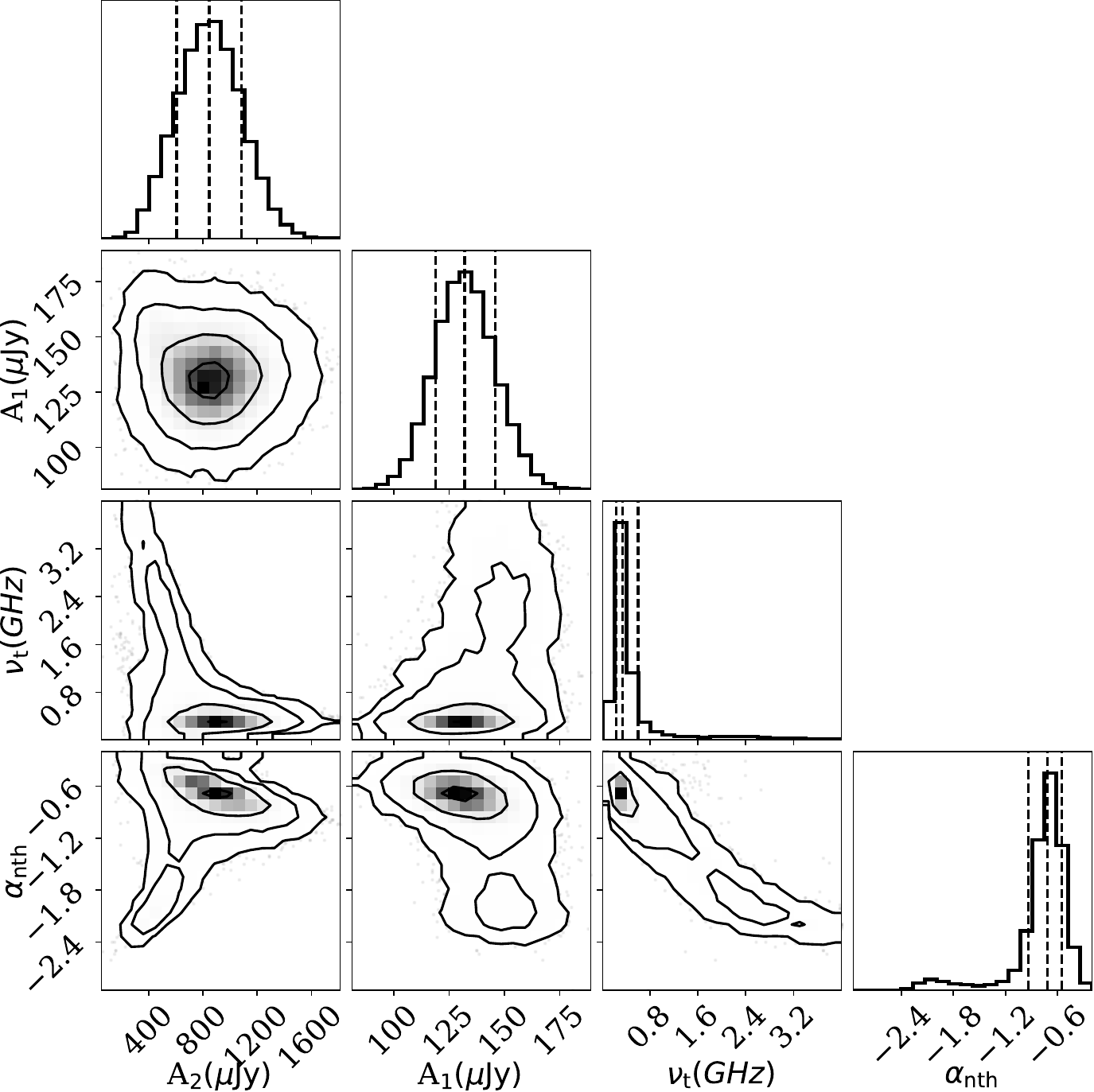 }}}%
  \caption{\revtext{ Bayesian \rcsed{} fitting results for \bbten{} using the \revtexts{SFG-thick} model. The various panels follow the same order as Figure \ref{fig: J1032 bayes rado-sed fit} in terms of the choice of priors on \sth{}. We constrain the \nuturn{} to $\sim 0.34 - 0.39$ GHz across the different choice of priors (Table \ref{table: bayes parameters}). The \rcsed{} is well fit with a non-thermally dominant \revtexts{SFG-thick} \rcsed{} typical of normal star-forming galaxies and unlike other \xsfg{s} from our sample. See Section \ref{sec: radio-sed individual cases} for a detailed discussion on this source.} }

  \label{fig: BB10 bayes rado-sed fit}
\end{figure*}

\begin{figure*}
\captionsetup[subfigure]{labelformat=empty}
\centering
\subfloat[]{{\includegraphics[width=0.44\hsize]{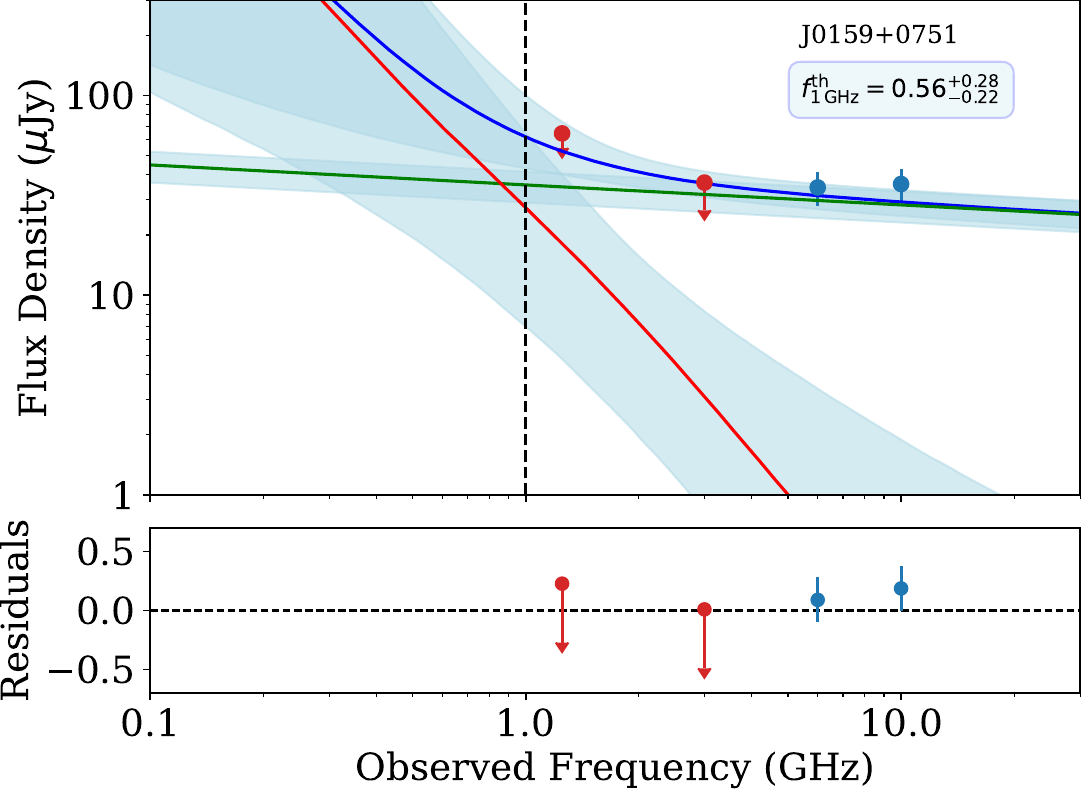} }}%
\qquad
\subfloat[]{{\includegraphics[width=0.44\hsize]{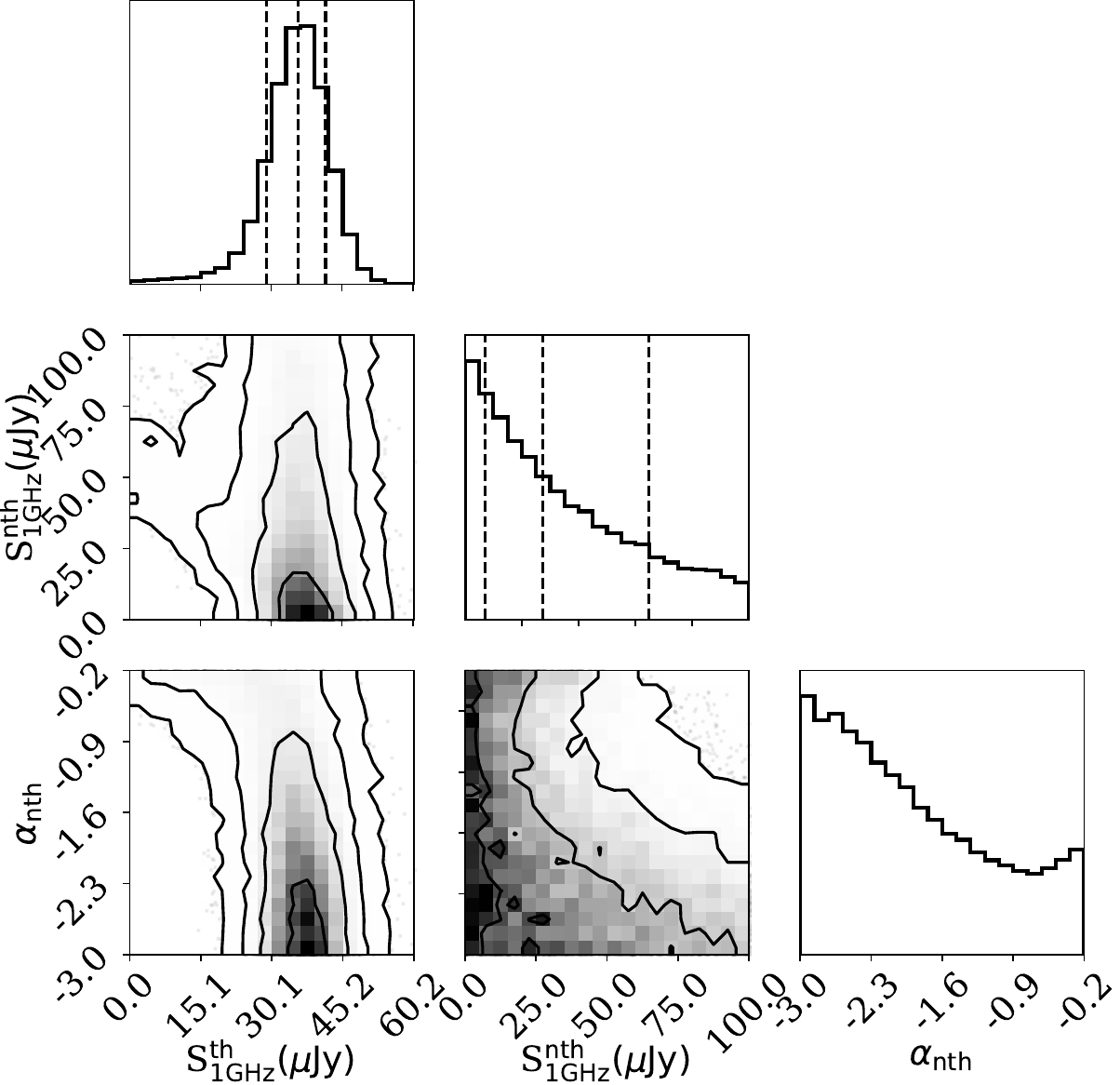 }}}%
  \qquad
\subfloat[]{{\includegraphics[width=0.44\hsize]{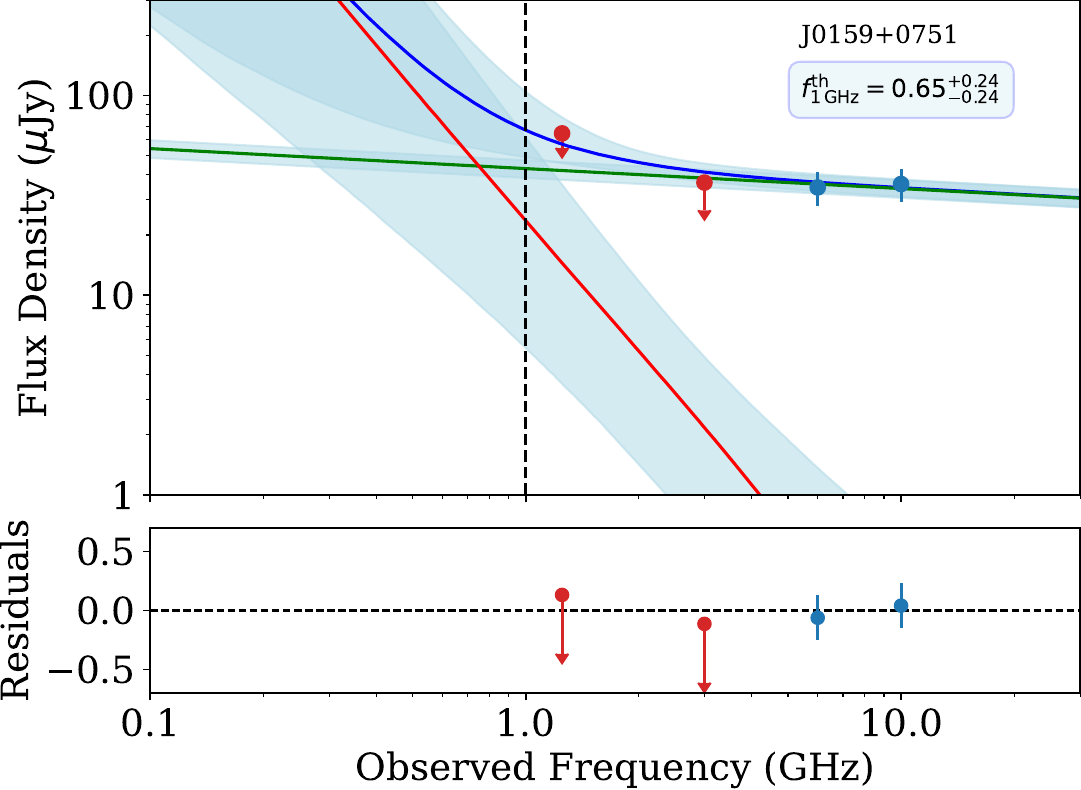} }}%
\qquad
  \subfloat[]{{\includegraphics[width=0.44\hsize]{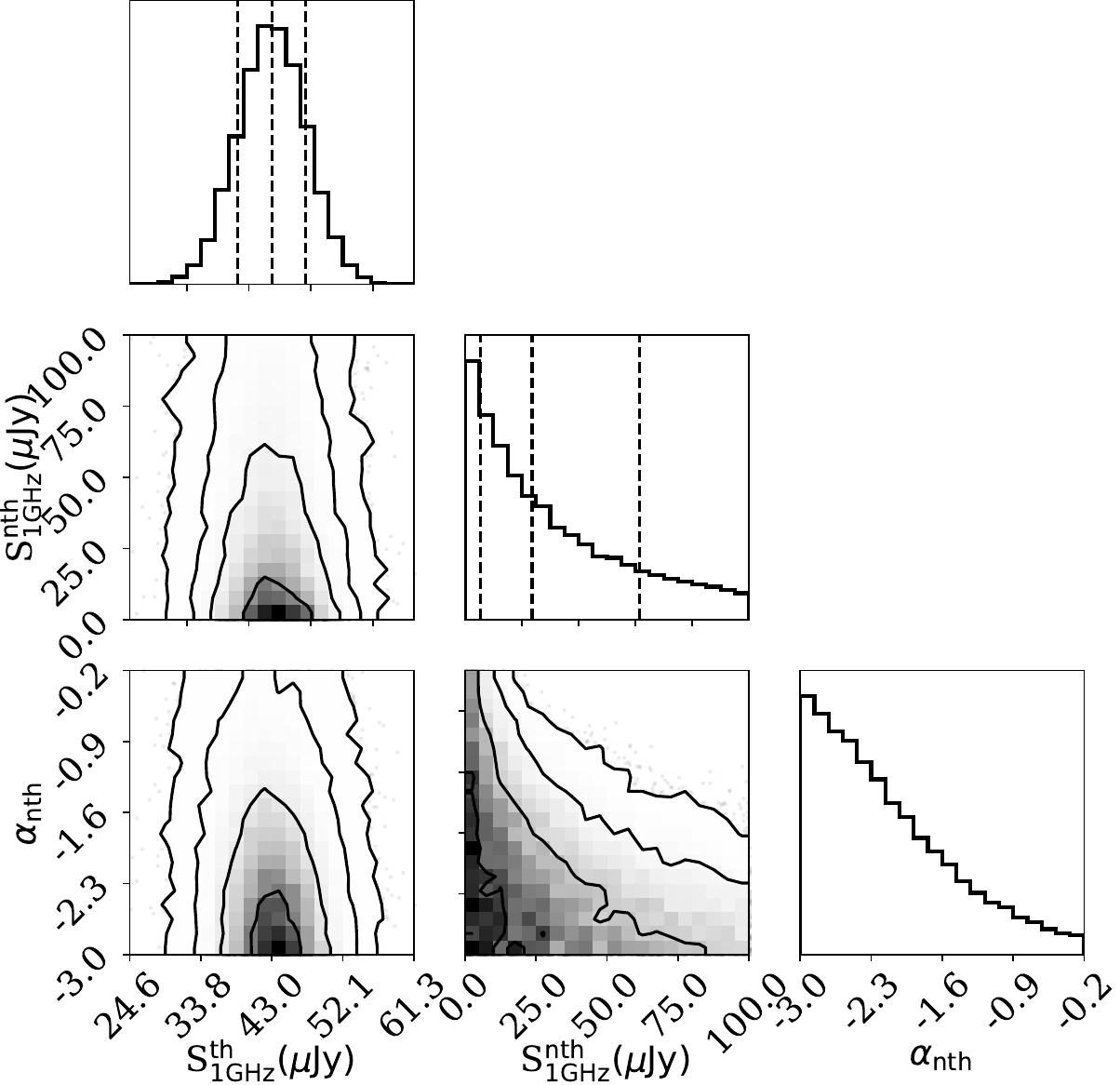 }}}%
  \caption{\revtext{ Bayesian \rcsed{} fitting results for J0159+4919 using the \revtexts{SFG-thin} model. The various panels follow the same order followed in Figure \ref{fig: BB13 bayes rado-sed fit}. See Section \ref{sec: radio-sed individual cases} and Table \ref{table: bayes fitting} for details on the priors, choice of model and the bands used for the Bayesian fitting. Table \ref{table: bayes parameters} summarizes the constraints on the various parameters and constraints on \fth{}. We find that the \rcsed{} is well fit with a thermally dominant component.} }

  \label{fig: J0159 bayes rado-sed fit}
\end{figure*}

\begin{figure*}
\captionsetup[subfigure]{labelformat=empty}
\centering
\subfloat[]{{\includegraphics[width=0.44\hsize]{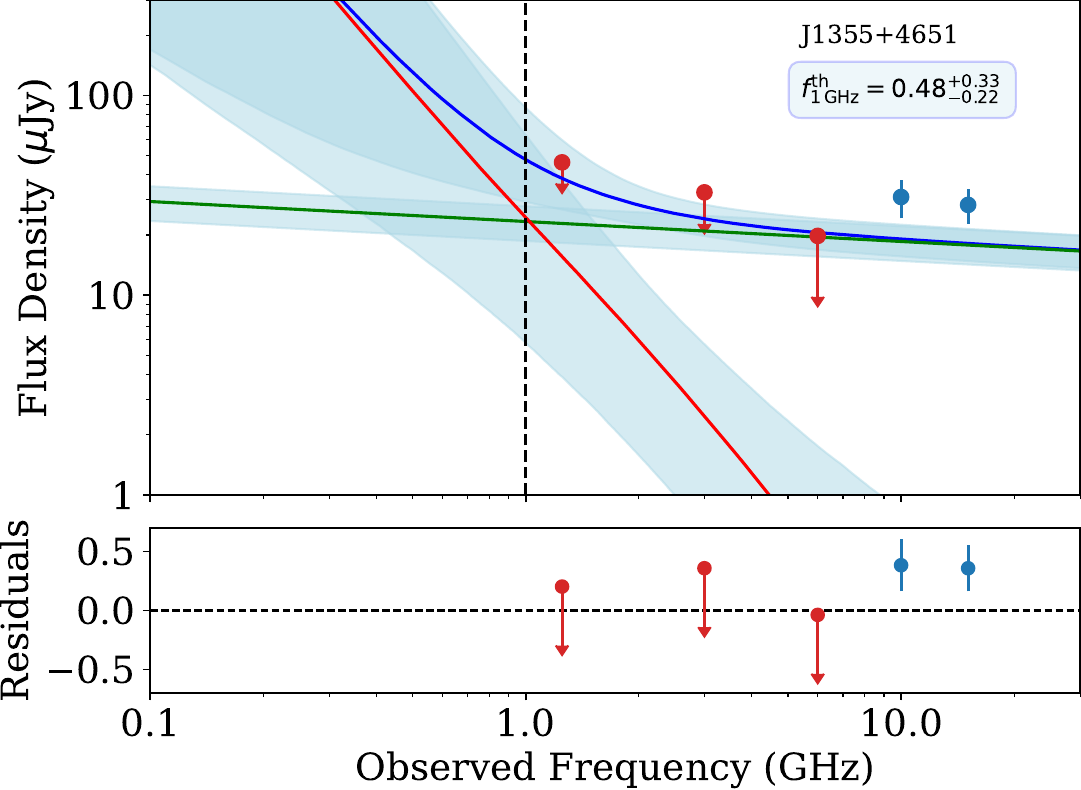} }}%
\qquad
\subfloat[]{{\includegraphics[width=0.44\hsize]{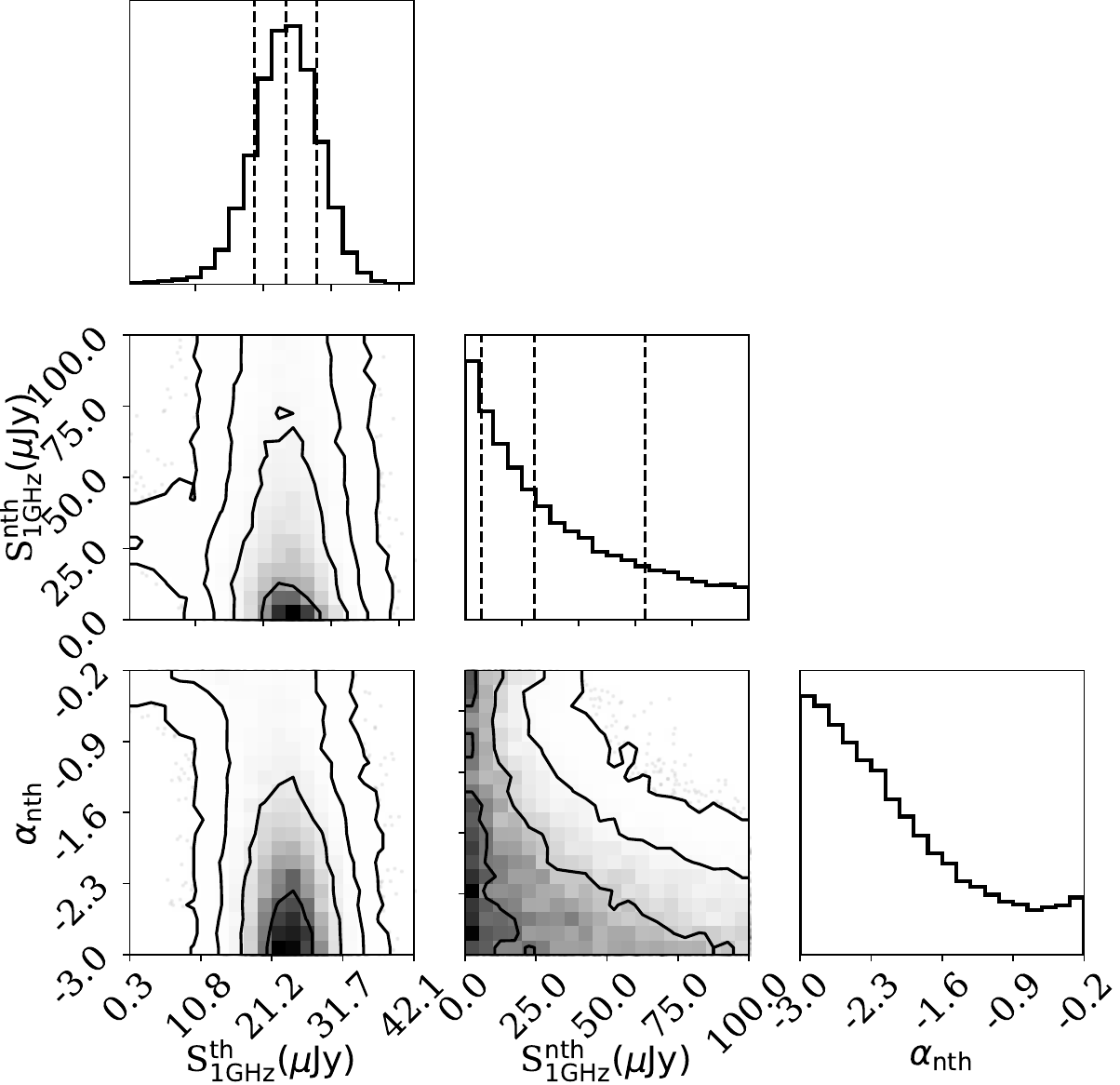 }}}%
  \qquad
\subfloat[]{{\includegraphics[width=0.44\hsize]{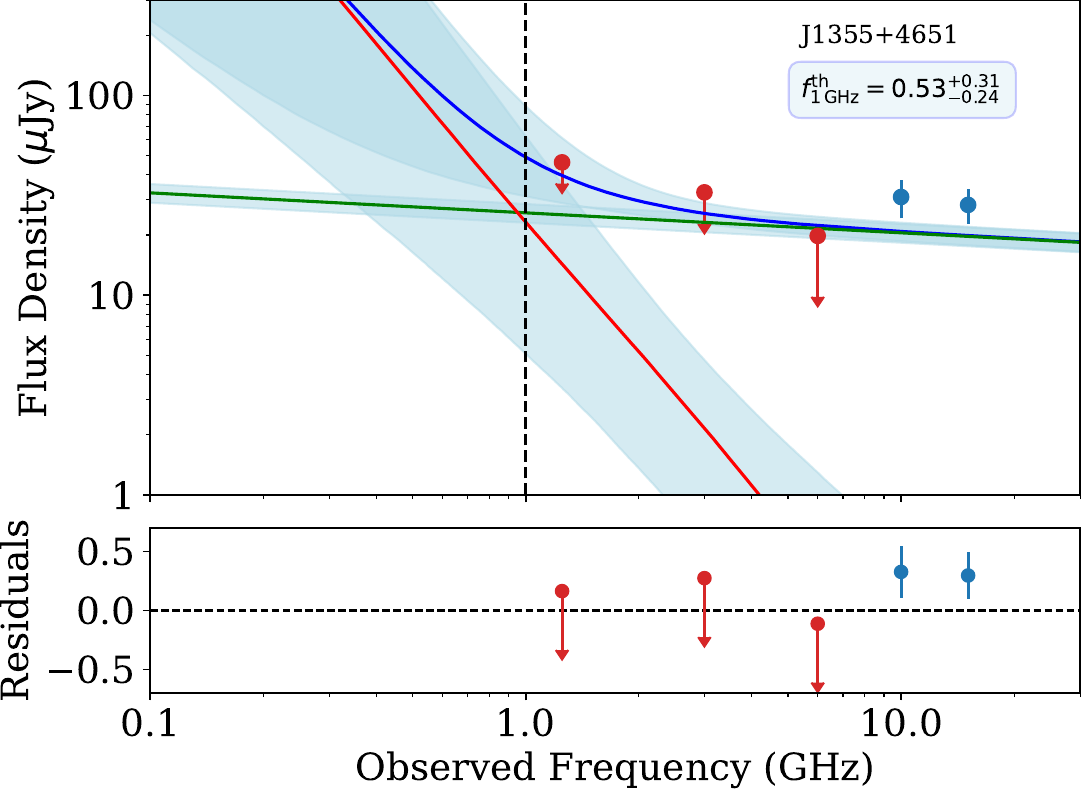} }}%
\qquad
  \subfloat[]{{\includegraphics[width=0.44\hsize]{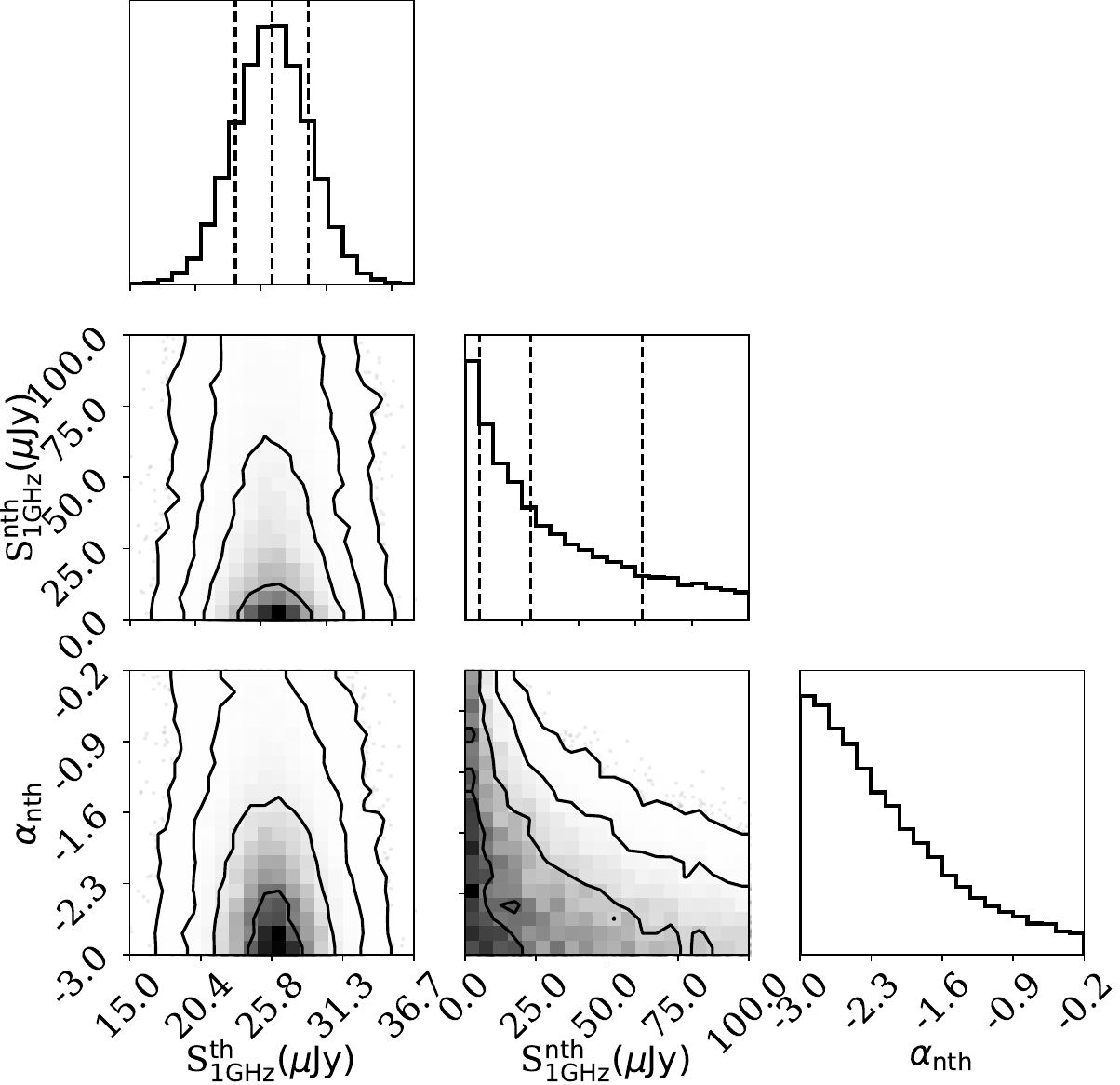 }}}%
  
  \caption{\revtext{ Bayesian \rcsed{} fitting results for J1355+4651 using the \revtexts{SFG-thin} model. The various panels follow the same order as Figure \ref{fig: BB13 bayes rado-sed fit} in terms of the choice of priors on \sth{}. Overall we find that the \rcsed{} is well fit with a thermally dominant \revtexts{SFG-thin} \rcsed{}. See Section \ref{sec: radio-sed individual cases} for a detailed discussion on this source. We notice weak systematic offset ($<2\sigma$) in the residuals at $10.0$ and $15.0$ GHz bands.} }

  \label{fig: J1355 bayes rado-sed fit}
\end{figure*}

\revtexts{We find that the observed \rcsed{s} shown as blue points and red arrows in Figures \ref{fig: BB13 bayes rado-sed fit} - \ref{fig: J1355 bayes rado-sed fit} have a flat spectral index between 6-15 GHz for most sources, with some sources (J1032+4919 and \bbten{}) showing evidence for a turnover at lower frequencies.} \bbten{} is the only source in our sample which shows a fairly steep ($\alpha \approx -1.0$) below 6 GHz to up to $\sim 1$ GHz. Between 1 GHz - 150 MHz range it shows evidence for a turnover (or flattening). The presence of a strong turnover at lower frequencies in the radio spectrum of star-forming galaxies can be a signature of free-free absorption (FFA)\footnote{Synchrotron self-absorption (SSA) can also lead to low-frequency turnovers. However, the brightness temperature (T$_b$) of our sources is too small ($\leq 100$ K) for SSA to be significant at GHz frequency range, which will require an extremely high T$_b \approx 10^{12}$ K \citep{Condon_16_essential_RA}.}. The radio spectrum can show such a turnover at GHz frequencies due to a high emission measure (EM), owing to high ionised gas density \citepalias{Hunt04}. Such physical conditions are typical of \xsfg{s} \citep{Izotov20-diverseLyA, Izotov24-LyA-metalpoor}. \sbs{}, a nearby low-metallicity starburst, also shows such a turnover but at a slightly lower frequency of $\sim 1.5$ GHz, although at higher frequencies it is relatively steeper than in xSFGs from our sample \citepalias{Hunt04}. In general, the radio spectra can be flattened due to a change in the dominant cosmic-ray cooling \citep{Hummel91, Chyzy18}, however it cannot explain a complete turnover (positive spectral index) in the radio spectra.

\subsection{Radio-SED Models}
\label{sec: radio-sed models}

We particularly, consider two \rcsed{} models,
\begin{itemize}
    \item Star-forming galaxy (SFG)-thin:\ Here the total radio emission ($S_{\nu}$) is optically thin and comprises of a mix of a single thermal ($S_{\nu}^{\mathrm{th}}$) and non-thermal component ($S_{\nu}^{\mathrm{nth}}$), described as:

\begin{equation}\label{eq: simple sed model}
    S_{\nu} = S_{\nu}^{\mathrm{th}} + S_{\nu}^{\mathrm{nth}} 
    = A_1\left(\frac{\nu}{\nu_\circ}\right)^{-0.1} + A_2 \left(\frac{\nu}{\nu_\circ}\right)^{\alpha_\mathrm{nth}}.
\end{equation}

\revtext{Here, $\nu$ is the observed frequency, $\nu_\circ$ is a reference frequency which is fixed to 1 GHz in all our models, thus $A_1 = \sth$, and $A_2 = \snth$. And \alphanth{}\footnote{In the following, we use this sign convention to define the spectral index. Note that some references use the opposite sign convention.} is the non-thermal spectral index. We then define the thermal fraction derived from the \rcsed{} at a reference frequency of 1 GHz (\fth{}) as $\fth{} = S_{1 \mathrm{GHz}}^{\mathrm{th}}/S_{1 \mathrm{GHz}}$. In the most general case there are three free parameters: a) \sth, b) \snth and c) \alphanth{}. }

\item  \revtexts{ \revtexts{SFG-thick} Model}:\ \revtexts{Here the \rcsed{} is optically thick and thus shows an additional strong turnover at lower frequencies due to the effect of FFA arising due to high EM in the ISM. Following \citet{Galvin18, Clemens10, condon1992}, we describe the \revtexts{SFG-thick} SED as:}
\begin{equation}\label{eq: FFA model}
S_{\nu} = \left( 1 - \exp\left(-\tau_\nu^{\mathrm{ff}}\right) \right)
\left( A_1 + A_2 \left(\frac{\nu}{\nuturn{}}\right)^{0.1 + \alphanth{}} \right) \left(\frac{\nu}{\nuturn{}}\right)^{2}
\end{equation}
\revtext{Here the optical depth (\odepth{}) is parametrized using the turnover frequency (\nuturn{}) as $\tau_{\nu}^{\mathrm{ff}} = (\nu/\nuturn{})^{-2.1}$. Thus, \nuturn{} is the frequency at which the optical depth becomes 1 \citep[see e.g.,][]{Hunt04, Clemens10, Galvin18}. In the most general case the free parameters in this model are: a) $A_1$, b) $A_2$, c) \nuturn{} and d) \alphanth{}. }\revtext{By constraining \nuturn{}, we can then derive the EM as it is directly related to \odepth{}, such that increasing the EM leads to a higher value of \odepth{} (and also of \nuturn{}). Specifically, in this work we used the relation from \citetalias{Hunt04} (their Eq. 1) to derive the EM.}

\end{itemize}

\subsection{Bayesian model fitting procedure}
\label{sec: radio-sed fitting}
\revtext{We incorporate a Bayesian model fitting approach to constrain the thermal/non-thermal emission and to determine the presence of FFA component. We constrain the free parameters in our \rcsed{s} using Markov Chain Monte Carlo (MCMC) using the \emcee{} package \citep{Foreman-Mackey13}. We use all the bands in which we have obtained the flux densities together with their observational uncertainities. We also incorporate upper limits on those bands in which we have non-detections. Thus our likelihood function has a mixed form consisting of a term for bands with detection and those with non-detections using the approach derived in \citet{Sawicki12}. This approach is routinely applied in popular SED fitting tools (e.g., \textsc{CIGALE}) to properly treat upper limits \citep[][]{Boquien19, MRMOOSE18}. }

\revtext{We thus construct our log likelihood function, following the derivation in \citet[][]{Sawicki12} and \citet{MRMOOSE18}, as}

\begin{align}
\ln \mathcal{L}(\theta) =\; &
\sum_{\nu_{\mathrm{det}}} 
\left( \frac{S^{\mathrm{obs}}_{\nu_{\mathrm{det}}} - S^{\mathrm{model}}_{\nu_{\mathrm{det}}}(\theta)}{\sigma_{\nu_{\mathrm{det}}}} \right)^{2} \notag \\
& - 2 \sum_{\nu_{\mathrm{nondet}}} \ln \left\{
\sqrt{\frac{\pi}{2}}\, \sigma_{\nu_{\mathrm{nondet}}}
\left[
1 + \operatorname{erf} \left(
\frac{S^{\mathrm{lim}}_{\nu_{\mathrm{nondet}}} - S^{\mathrm{model}}_{\nu_{\mathrm{nondet}}}(\theta)}{\sqrt{2}\, \sigma_{\nu_{\mathrm{nondet}}}}
\right)
\right]
\right\}.
\end{align}
\revtext{Here S$^{\mathrm{obs}}_\nu$ and S$^{\mathrm{model}}_\nu$ represent the observed and model flux density respectively at a fixed band with detection ($\nu_{\mathrm{det}}$) and non-detection ($\nu_{\mathrm{nondet}}$). The noise in the detected and non-detected bands is represented by $\sigma_\mathrm{det}$ and $\sigma_\mathrm{nondet}$ respectively. The free parameters in the model are represented by $\theta$. The S$^{\mathrm{lim}}_{\nu_{\mathrm{nondet}}}$ in the error function ($\operatorname{erf}$) represents the value of the upper limit on the flux density generally set to 1$\sigma$ \citep{Sawicki12}. For all our model fits, we run the MCMC with 50 walkers for 10,000 steps. We discard the first 200 steps for burn-in and thin them by a factor of 15. Table \ref{table: bayes fitting} provides additional information on the Bayesian fitting for individual sources, particularly the choice of priors on the parameters and the radio bands used for fitting.}

\subsubsection{Choice of priors} 
\label{sec: priors}
We adopt \revtexts{uniform priors} on \snth{} in a wide range for all our fits (see Table \ref{table: bayes fitting}) \revtexts{with the lower bound close to zero to maintain positivity.} For \sth{} we show the fitting results with \revtexts{uniform priors} in the same range as \snth{} and also with Gaussian priors. The centre and widths of the Gaussian priors on \sth{} were adopted from the physically motivated dust-corrected \hbeta{} flux density. In Appendix \ref{appendix sec: radio-based Hbeta flux details} we describe our procedure of relating the dust-corrected \hbeta{} flux density and \sth{}. Further, we provide \revtexts{uniform priors} on \alphanth{} for all our fits in a wide range of $-3.0 < \alphanth{} < -0.2 $. The choice of extending the prior on \alphanth{} to a very flat value of $-0.2$ is because in star-forming galaxies \alphanth{} is found to flatten with increasing \sfrdensity{} and sSFR \citep[e.g.,][]{tabatabaei2017, Tabatabaei25, Basu15}. The flattening of \alphanth{} is due to a more energetic population of CRs in starburst galaxies. \xsfg{s} are known to posses high \sfrdensity{} and sSFR and thus we extend the range of the prior on \alphanth{} to such a flat value.  Additionally, in compact starburst galaxies with a high ISM  gas density ionization and bremsstrahlung losses can be significant and flatter \alphanth{} at GHz frequencies than that expected from merely synchrotron losses \citep{Thompson06, Murphy09, Lacki10}. In the two cases, where we attempt a modelling with \revtexts{SFG-thick} model, we also provide \revtexts{uniform priors} on \nuturn{} in a wide range.

\revtext{The details on the Bayesian modelling with different priors and the radio bands used for fitting is summarized in Table \ref{table: bayes fitting}. }

\subsection{Modelling individual cases}
\label{sec: radio-sed individual cases}

\begin{table*}
\caption{\revtexts{Details on priors, radio bands and models used for the Bayesian model fitting for each of the source in our sample.}}
\label{table: bayes fitting}

\renewcommand{\arraystretch}{1.2} 
\setlength{\tabcolsep}{6pt} 
\begin{center}

\begin{tabular}{c c p{7.0cm} c}
\hline
Source & \rcsed{} Model & Free parameters \& Priors & Bands (used for fitting)  \\
\hline \hline
\multirow{2}{*}{\bbthirteen{}} 
& \multirow{2}{*}{\revtexts{SFG-thin}}  & \sth{}, \snth{}: [$10^{-6}$, 100] $\mu$Jy (\revtexts{uniform}), \alphanth{}: [$-3.0$, $-0.2$] (\revtexts{uniform}) & \multirow{2}{*}{\lofar{}, \gmrt{} (Band-5), \vla{} (C, X, Ku)} \\
&   & \sth{} (Gaussian: $50.8 \pm 5.1$ $\mu$Jy), \snth{}: [$10^{-6}$, 100] $\mu$Jy (\revtexts{uniform}), \alphanth{}: [$-3.0$, $-0.2$] (\revtexts{uniform}) & \\
\hline
\multirow{2}{*}{J1032+4919} 
& \multirow{2}{*}{\revtexts{SFG-thick}}  & \sth{}, \snth{}: [$10^{-6}$, 100] $\mu$Jy (\revtexts{uniform}), \nuturn{}: [\revtexts{1}, 7] GHz (\revtexts{uniform}), \alphanth{}: [$-3.0$, $-0.2$] (\revtexts{uniform}) & \multirow{2}{*}{\gmrt{} (Band-5), \vla{} (L, S, C, X, Ku)} \\
&   & \sth{} (Gaussian: $57.0 \pm 7.4$ $\mu$Jy), \snth{}: [$10^{-6}$, 100] $\mu$Jy (\revtexts{uniform}), \nuturn{}: [\revtexts{1}, 7] GHz (\revtexts{uniform}), \alphanth{}: [$-3.0$, $-0.2$] (\revtexts{uniform}) & \\
\hline
\multirow{2}{*}{\bbten{}} 
& \multirow{2}{*}{\revtexts{SFG-thick}}  & \sth{}, \snth{}: [1, 2000] $\mu$Jy (\revtexts{uniform}), \nuturn{}: [0.1, 7] GHz (\revtexts{uniform}), \alphanth{}: [$-3.0$, $-0.2$] (\revtexts{uniform}) & \multirow{2}{*}{\lofar{}, \gmrt{} (Band-5), \vla{} (S, C, X, Ku)} \\
&   & \sth{} (Gaussian: $126.3 \pm 12.6$ $\mu$Jy), \snth{}: [1, 2000] $\mu$Jy (\revtexts{uniform}), \nuturn{}: [0.1, 7] GHz (\revtexts{uniform}), \alphanth{}: [$-3.0$, $-0.2$] (\revtexts{uniform}) & \\ 
\hline
\multirow{2}{*}{J0159+0751} 
& \multirow{2}{*}{\revtexts{SFG-thin}}  & \sth{}, \snth{}: [$10^{-6}$, 100] $\mu$Jy (\revtexts{uniform}), \alphanth{}: [$-3.0$, $-0.2$] (\revtexts{uniform}) & \multirow{2}{*}{\gmrt{} (Band-5), \vla{} (S, C, X)} \\
&   & \sth{} (Gaussian: $53.1 \pm 7.0$ $\mu$Jy), \snth{}: [$10^{-6}$, 100] $\mu$Jy (\revtexts{uniform}), \alphanth{}: [$-3.0$, $-0.2$] (\revtexts{uniform}) & \\
\hline
\multirow{2}{*}{J1355+4651} 
& \multirow{2}{*}{\revtexts{SFG-thin}}  & \sth{}, \snth{}: [$10^{-6}$, 100] $\mu$Jy (\revtexts{uniform}), \alphanth{}: [$-3.0$, $-0.2$] (\revtexts{uniform}) & \multirow{2}{*}{\gmrt{} (Band-5), \vla{} (S, C, X, Ku)} \\
&   & \sth{} (Gaussian: $27.3 \pm 3.6$ $\mu$Jy), \snth{}: [$10^{-6}$, 100] $\mu$Jy (\revtexts{uniform}), \alphanth{}: [$-3.0$, $-0.2$] (\revtexts{uniform}) & \\
\hline
\end{tabular}

\end{center}
\end{table*}



\begin{table*}
\centering
\caption{\revtexts{Constrained values of the parameters based on our Bayesian radio SED fitting.}}
\label{table: bayes parameters}
\begin{tabular}{c c c c c c c}
\hline
Source & \rcsed{} Model & \sth{} ($\mu$Jy) & \snth{} ($\mu$Jy) & \alphanth{}$^{a}$ & $\nu_{t}$ (GHz) & $f^{\text{th (SED)}}_{1\text{GHz}}$ \\
\hline \hline

\multirow{2}{*}{J160810+352809} & SFG-thin  & $53.57^{+5.14}_{-7.33}$ & $22.19^{+23.81}_{-15.51}$ & $--$ & $--$ & $0.70^{+0.19}_{-0.19}$ \\
& SFG-thin (Gaussian prior on \sth{}) & $52.62^{+3.48}_{-3.79}$ & $21.20^{+22.41}_{-14.31}$ & $--$ & $--$ & $0.71^{+0.18}_{-0.17}$ \\
\hline

\multirow{2}{*}{J1032+4919} & SFG-thick & $66.62^{+8.90}_{-20.55}$ & $21.97^{+28.04}_{-16.15}$ & $--$ & $3.36^{+0.92}_{-0.65}$ & $0.51^{+0.36}_{-0.29}$ \\
& SFG-thick (Gaussian prior on \sth{}) & $60.73^{+6.98}_{-7.35}$ & $29.30^{+14.97}_{-14.62}$ & $--$ & $3.45^{+1.02}_{-0.69}$ & $0.49^{+0.26}_{-0.26}$ \\
\hline

\multirow{2}{*}{J150934+373146} & SFG-thick & $217.44^{+20.86}_{-22.64}$ & $622.54^{+306.87}_{-202.30}$ & $-1.79^{+0.47}_{-0.31}$ & $0.97^{+0.73}_{-0.45}$ & $0.25^{+0.09}_{-0.10}$ \\
& SFG-thick (Gaussian prior on \sth{}) & $131.97^{+13.92}_{-13.23}$ & $847.02^{+237.20}_{-240.09}$ & $-0.71^{+0.16}_{-0.22}$ & $0.34^{+0.26}_{-0.10}$ & $0.19^{+0.03}_{-0.03}$ \\
\hline

\multirow{2}{*}{J0159+0751} & SFG-thin  & $35.62^{+5.90}_{-6.56}$ & $27.42^{+37.37}_{-20.45}$ & $--$ & $--$ & $0.56^{+0.28}_{-0.22}$ \\
& SFG-thin (Gaussian prior on \sth{}) & $43.00^{+4.37}_{-4.40}$ & $23.63^{+37.90}_{-18.19}$ & $--$ & $--$ & $0.65^{+0.24}_{-0.24}$ \\
\hline

\multirow{2}{*}{J1355+4651} & SFG-thin  & $23.37^{+4.49}_{-4.70}$ & $24.49^{+38.95}_{-18.72}$ & $--$ & $--$ & $0.48^{+0.33}_{-0.22}$ \\
& SFG-thin (Gaussian prior on \sth{}) & $25.82^{+2.77}_{-2.79}$ &  $23.25^{+39.25}_{-18.16}$ & $--$ & $--$ & $0.53^{+0.31}_{-0.24}$ \\
\hline

\end{tabular} \\
\hbox{Notes. }
\hbox{\revtexts{Each column reports the median parameter value and the corresponding 16th–84th percentile confidence interval.}}
\hbox{$^a$ For sources with empty values we are unable to contrain the \alphanth{} due to lack of detections at low-frequencies.}

\end{table*}

\revtext{Here we model the individual sources with different \rcsed{} using a Bayesian approach which allows us to investigate the various degeneracies in the parameters and robustly estimate the thermal/non-thermal component and the dervied physical parameters (e.g., \fth{} and EM).}

\subsubsection{\bbthirteen{}}
\revtext{
As seen in Figure \ref{fig: BB13 bayes rado-sed fit}, this source has a flat radio spectrum from $1.2 - 15$ GHz, and can thus be modelled using a \revtexts{SFG-thin} \rcsed{}. We have detections in $4$ bands and a non-detection in one \lofar{} band for this source. We show the results of the Bayesian fitting using two combinations of priors on \sth{}:(i) \revtexts{uniform prior} in the range of $10^{-6} - 100 ~\mu$Jy (see top row in Figure \ref{fig: BB13 bayes rado-sed fit}) and (ii) Gaussian priors with the centre and width adopted from the dust-corrected \hbeta{} flux density (see bottom row in Figure \ref{fig: BB13 bayes rado-sed fit}).  The left column in Figure \ref{fig: BB13 bayes rado-sed fit} shows the corresponding best-fit median (50th percentile) \revtexts{SFG-thin} model in solid blue line. The individual contribution from the median thermal and non-thermal emission are shown in solid green and red line, respectively. The shaded region shows the 16th and 84th percentile range from the MCMC posterior samples. We also show the fractional residuals between the observed data and the median model for bands with detections and with red upper limits for non-detections in the left panels. The right column in Figure \ref{fig: BB13 bayes rado-sed fit} shows the posteriors for \sth{}, \snth{} and \alphanth{} and the corresponding 1D marginalized posterior distribution for the two cases of priors. \revtexts{Here the dashed vertical lines show the 16th, 50th (median) and 84th percentiles for the 1D marginalized posterior.} See Table \ref{table: bayes fitting} for the different choices on prior. We find that the \revtexts{SFG-thin} model fits well with the observed data with some degeneracy between \sth{} and \snth{}. Using Gaussian priors reduces the degeneracy between these parameters. In both cases due to a lack of detections at low-frequencies ($< 1$GHz) we are unable to constrain \alphanth{}. The median, 16th and 84th percentile values of the fitted parameters for the two cases of priors are shown in Table \ref{table: bayes parameters}. } 

Finally, we derive the \fth{} and the uncertainities on it based on the MCMC posterior samples. The median (best-fit) value of \fth{} is high ($\sim 0.7$) in both cases (Table \ref{table: bayes parameters}). The uncertainty (1$\sigma$) on \fth{} is in the range of $0.2$ in both cases. Thus overall our Bayesian analysis supports a thermally dominant \revtexts{SFG-thin} \rcsed{} for \bbthirteen{}. \revtexts{To ensure that our constraints on the non-thermal component were not biased by the lower bound on the \snth{} prior (\snth{} $\gtrsim 10^{-6}\mu$Jy), we performed a statistical validation by re-running the fit with a uniform prior allowing for negative flux values for \snth{}, knowing that this spans the unphysical range for the flux densities. The resulting posterior for \snth{} was symmetrically centered on zero, thus robustly confirming the non-negligible contribution of the non-thermal component. This validation demonstrates that our original prior does not introduce significant bias and confirms the robustness of the derived thermal dominance.}

\subsubsection{J1032+4919}
\revtext{As seen in Figure \ref{fig: J1032 bayes rado-sed fit}, this source has a flat spectrum at high frequency and the upper limits ($3\sigma$) at lower-frequencies suggest a turnover in the \rcsed{}. The turnover is observed in the \gmrt{} Band-5 ($1.2$ GHz) and the VLA L-band ($1.5$ GHz) both of which are independently observed using two different radio telescopes. A similar level of upper limit was found by \citet{Sebastian19} in their independent \gmrt{} observation (proposal id: 34\_123, PI: Sebastian) of this source \footnote{Additionally the \gmrt{} data in their study and presented here is reduced using two different radio data reduction pipelines and was based on \aips{} in their study and CASA here.}. Additionally the upper limits shown in red in Figure \ref{fig: J1032 bayes rado-sed fit} represent $3\sigma$ levels thus increasing the robustness of the turnover. We thus fit this source with the \revtexts{SFG-thick} model to better model the turnover in the \rcsed{}.  We follow a similar procedure to the one followed for \bbthirteen{}, and present results of the Bayesian fitting with two different values priors on \sth{} shown in Figure \ref{fig: J1032 bayes rado-sed fit}. We let the \nuturn{} to have a \revtexts{uniform prior} between $\revtexts{1} - 7$ GHz. See Table \ref{table: bayes fitting} for details.}

\revtext{We find that the data is well fit with a \revtexts{SFG-thick} model for both the choice of priors. In the case of \revtexts{uniform priors} on all the parameters there is a strong degeneracy between the thermal, non-thermal component and \nuturn{}. \revtexts{Note that, strictly speaking it appears that the likelihood on the non-thermal component and \alphanth{} favours a range beyond our chosen priors. However by choosing hard physical priors we exclude this region of the parameter space due to their unphysical nature (see \ref{sec: priors}).} Using Gaussian priors on \sth{} informed from the dust-corrected \hbeta{} flux density helps to reduce this degeneracy. \revtexts{Additionally, the posterior distribution on the non-thermal component is largely within our choice of prior.} As shown in Table \ref{table: bayes parameters}, we are able to well constrain the \nuturn{} to $\sim 3.4$~GHz (median value) and is consistent within the two choice of priors on \sth{}. The estimated EM of this source is relatively high $\sim (9 - 10) \times 10^{7}$ pc cm$^{-6}$ (see Table \ref{table: physical properties}). Similar to the previous case, \alphanth{} remains unconstrained due to lack of detections at low-frequencies. In spite of this, our Bayesian \rcsed{} analysis favours a thermally dominant SED with the median value (best-fit) of \fth{} in the range of $\sim 0.5$ for both the choice of priors on \sth{}.}

\subsubsection{\bbten{}}
\bbten{} has a \rcsed{} below $10$ GHz which is consistent with a steep spectrum and a flattens below $\sim 1$GHz between the \gmrt{} Band-5 and \lofar{} band. Such a flattening can occur due to the effect of FFA. We thus fit the observed \rcsed{} with the \revtexts{SFG-thick} model following the similar approach as followed for J1032+4919. We choose \revtexts{uniform priors} in a wide range for \nuturn{}, \snth{} and \alphanth{} (see Table \ref{table: bayes fitting}). In Figure \ref{fig: BB10 bayes rado-sed fit} we show the results with flat (top row) and Gaussian (bottom row) priors on \sth{}. We find that the \rcsed{} is non-thermally dominant for both the choice of priors, as opposed to the other objects studied in our sample, consistent with its steep apparent spectral index. The median value of \fth{} is close to $0.2 - 0.3$ close to that typically found in normal star-forming galaxies in the nearby Universe \citep[e.g,][]{tabatabaei2017}. For this source we are able to constrain the \alphanth{}, however there is a degeneracy most strongly between \alphanth{}, the non-thermal component and \nuturn{}. In the case of \revtexts{uniform priors} on all parameters, our Bayesian model favours a steep \alphanth{} of $\sim -1.8$ and a \nuturn{} close to $1$GHz (and an EM of $\sim 0.7 \times 10^{7}$ pc cm$^{-6}$). Using Gaussian priors on \sth{} from the dust-corrected \hbeta{} line, reduces these degeneracies with well constraints on the thermal and non-thermal component. In this case, we find that a \revtexts{SFG-thick} model fits well to the data with an \alphanth{} close to the canonical value of $-0.7$ within the uncertainties, a \nuturn{} at a lower frequency of $\sim 0.3$ GHz (EM $\sim 8 \times 10^{5}$ pc cm$^{-6}$) and \fth{} of $\sim 0.2$. Table \ref{table: bayes parameters} provides a fitted values of the paramters for the two choice of priors.

\subsection{J0159+0751}
As seen in Figure \ref{fig: J0159 bayes rado-sed fit}, the \rcsed{} is flat based on the detections at the two high-frequencies bands ($>5$GHz) and is consistent with a fairly flat spectrum taking into account the non-detections ($3 \sigma$ upper limits) in the rest of the two bands. We fit the source with a \revtexts{SFG-thin} model similar to the case of \bbthirteen{} (see also Table \ref{table: bayes fitting} for choice of priors). Here as well we favour a thermally dominant \rcsed{} (\fth{} $\sim \revtexts{0.6} $) in the case of both flat and Gaussian priors on \sth{}. Using Gaussian priors on \sth{} puts better constraints on the thermal component and somewhat reduces the uncertanities on \fth{}, thus supporting the thermally dominant \rcsed{}. The contribution of \snth{} (non-thermal component) remains below $\sim \revtexts{62}~\mu$Jy ($84$th percentile), however we are unable to put strong lower limits on it. In both the cases we are unable to well constrain \alphanth{}, although using Gaussian priors on \sth{} favours steeper values of \alphanth{} based on the 1D posteriors. \revtexts{Similar to the previous case of J1032+4919, it appears that the likelihood on the non-thermal component and \alphanth{} favours a range beyond our chosen priors, however we exclude this region on the basis of it being unphysical using hard physical priors. Deeper data at low frequencies will be crucial to better constrain both \alphanth{} and \snth{} and the nature of the \rcsed{}  at low-frequencies in this case.}

\subsection{J1355+4651}
As seen in Figure \ref{fig: J1355 bayes rado-sed fit}, the \rcsed{} is also flat at high-frequencies ($\geq 10$~GHz) and is also consistent with being flat given the 3$\sigma$ upper limits on the \gmrt{} Band-5 ($1.2$~GHz) and VLA S-band ($3$~GHz). The upper limit on VLA C-band ($6$~GHz) suggests that there might be a turnover in the spectrum. For simplicity and lack of strong evidence for a turnover based on the observed data, we attempt for fit this data using the simpler \revtexts{SFG-thin} model similar to J0159+0751. We find that the \rcsed{} is consistent with a thermally dominant SED with \fth{} in the range of $\sim \revtexts{0.5}$. As in the case of J0159+0751, we are unable to well constrain \snth{} and \alphanth{}, although while using Gaussian priors we prefer \snth $\leq 63 ~\mu$Jy ($84$th percentile) and steeper values of \alphanth{}. We notice that there is a very weak offset (< $2\sigma$) in the residuals at $10.0$ and $15.0$ GHz in fits with both the versions of the choice of priors on \sth{}. \revtexts{Similar to the previous case we reject the unphysical region favoured by the likelihood using hard physical priors.Deeper data is required both to increase the significance of detections and to detect in the lower frequencies ($< 10.0$~GHz) bands to better understand the nature of the \rcsed{} of this source.}

\begin{table*}
\begin{center}
\caption{Physical properties of \xsfg{s} from our RC observations.}
\label{table: physical properties}
\begin{tabular}{cccccc} 
\hline \\
Target & \revtext{EM}$^a$ & $N_e^b$ & \hbetaradio{}$^c$ & \specindex$^d$ & \fesc{} (indirect)$^e$ \\ 
\hline \hline \\
J0159+0751 & $--$ & $ 660 \pm 300$ & $\revtext{61.72}$ $\pm$ $ \revtext{14.73} $ & $0.078^\mathrm{C}_\mathrm{X} \pm \revtext{0.528}$ & $0.091 \pm 0.03$ \\
 & & & & & \\
J1032+4919 &  $9.60^{+6.36}_{-3.49}(10.15^{+7.34}_{-3.80})$  & $560 \pm 50$   & $\revtext{111.38}$ $\pm$ $ \revtext{34.33} $ & $-0.150^\mathrm{S}_\mathrm{C} \pm \revtext{0.262}$ & $--$ \\
 & & & & & \\
J1355+4651 & $--$ & $1200 \pm 150$ & $\revtext{41.47}$ $\pm$ $ \revtext{9.70} $ & $-0.222^\mathrm{X}_\mathrm{Ku} \pm \revtext{0.716}$ & $0.169 \pm 0.03$ \\
 & & & & & \\
\bbten{}   & $0.73^{+1.64}_{-0.53}(0.08^{+0.20}_{-0.04}) $  &                & $\revtext{412.96}$ $\pm$ $ \revtext{41.0} $ & $-0.958^\mathrm{S}_\mathrm{C} \pm \revtext{0.239}$ & $0.034 \pm 0.01$ \\
\bbthirteen{} & $--$ &                & $\revtext{88.70}$ $\pm$ $ \revtext{22.29} $ & $-0.128^\mathrm{C}_\mathrm{X} \pm \revtext{0.416}$ & $0.31 \pm 0.13$ \\ 
\hline
\end{tabular}

\vspace{0.5cm}

\hbox{Notes. $^a$In units of $10^7\times~$pc cm$^{-6}$. \revtext{Values in the bracket are with Gaussian priors on the \sth{}.} } 
\hbox{Here we use the T$_\mathrm{e}$ values from \citet{Izotov20-diverseLyA} and \citet{Izotov24-LyA-metalpoor} .}
\hbox{For \bbten{} and \bbthirteen{} we use a fixed value of $2\times 10^4$K.}
\hbox{$^b$In units of cm$^{-3}$. Values are taken from \citet{Izotov20-diverseLyA, Izotov24-LyA-metalpoor}.}
\hbox{$^c$Values shown are in units of $10^{-16}$ ergs/s/cm$^2$. \revtext{Derived using \sth{} from the \rcsed{} modelling.} } 
\hbox{$^d$The \vla{} bands used to derive \specindex{} is shown around each value.}
\hbox{$^e$\fesc{} values taken from \citet{Izotov20-diverseLyA, Izotov24-LyA-metalpoor} which are indirectly estimated using \vsep{} \citep{Izotov18b-vsep}.}
\end{center}
\end{table*}

\subsection{Comparison between H$\beta$ dust attenuation derived using Balmer lines and RC emission}
\label{sec: hbeta flux comparison}

\begin{figure}
  \resizebox{\hsize}
   {!}{\includegraphics{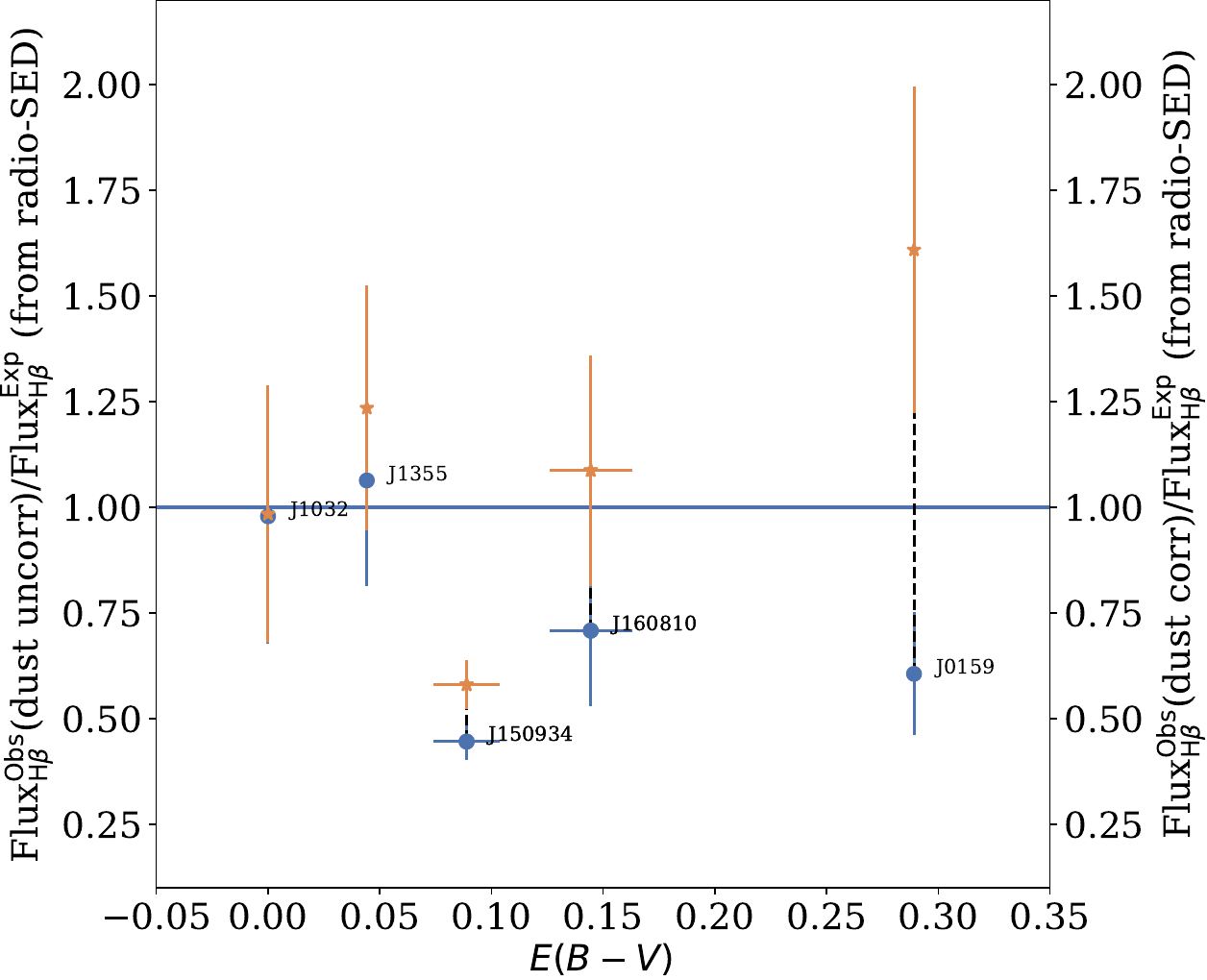}}
  \caption{\revtext{Left-axis: Ratio of the observed \hbeta{} flux density (dust uncorrected) to that expected from \sth{} (derived from the \rcsed{} assuming \revtexts{uniform priors}) vs. \ebv{} shown in circles. Right-axis: Same as left axis except that the observed \hbeta{} flux density is corrected for dust, shown in stars. The dashed line connects the points with and without dust correction. Overall we find that in spite of correcting the \hbeta{} flux density for dust extinction using hydrogen Balmer decrement, in a few cases (e.g., \bbten{}, \bbthirteen{} and J0159+0751) it is possible that a small fraction ($\sim20-30\%$) of the ionizing radiation emitted by the galaxy is not visible in the Balmer lines. See Sec. \ref{sec: hbeta flux comparison} for a detailed discussion and caveats.
  } }
  \label{fig: hbeta flux comparison}
\end{figure}

\revtext{The radio emission at GHz frequencies is not affected by dust extinction.  We can thus use the radio thermal (free-free) emission (\sth{}) estimated using the \rcsed{s} modelling, presented above, to independently derive the expected dust-free \hbeta{} flux density (\hbetaradio). Particularly, we use the \sth{} derived with \revtexts{uniform priors} which provides an independent estimate of the thermal emission using radio data alone. The free-free emission in a galaxy at both the radio and optical wavelengths arise from the same emitting regions. We can then use the ratio between the observed \hbeta{} (uncorrected for dust) and \hbetaradio{} to measure the total dust-extinction. In Appendix \ref{appendix sec: radio-based Hbeta flux details}, we describe the details on measuring \hbetaradio{} using \sth{}.} 
Table \ref{table: physical properties} shows the derived \hbetaradio{} for our sample. We use the observed and dust-extinction corrected \hbeta{} flux densities from \citet{Izotov20-diverseLyA} and \citet{Izotov24-LyA-metalpoor} that are derived from the SDSS spectrum. For \bbten{} and \bbthirteen{}, we measured the \hbeta{} line flux density from their SDSS spectrum. We estimated the dust extinction at \hbeta{} ($\mathrm{A}_{\mathrm{H}\beta}$) following the procedure described in \citet{Izotov16a}\footnote{Here the $\mathrm{A}_{\mathrm{H}\beta}$ is derived using $C$(\hbeta{}) which depends on the $\mathrm{A}_\mathrm{V}$ and $\mathrm{R}_{\mathrm{V}}$. Effectively this assumes the \citet{Cardelli89} extinction law which is found to be a better fit for such extreme galaxies \citep{Izotov17}. }, where we use the V-band extinction ($\mathrm{A}_\mathrm{V}$) from \citet{Jaskot19} and assume a $\mathrm{R}_{\mathrm{V} }$ of 2.7 \citep{Izotov17}.

In Figure \ref{fig: hbeta flux comparison} (left-axis), we compare this ratio against \ebv{}\footnote{Here \ebv{} is derived using the $C$(\hbeta{}) values from \citet{Izotov20-diverseLyA} and \citet{Izotov24-LyA-metalpoor}, and using $C$(\hbeta{}) $= 1.47\times$\ebv{}.} \revtext{shown in blue circles}. In the right-axis of Figure \ref{fig: hbeta flux comparison}, we compare the ratio between the dust-corrected \hbeta{} flux density, derived using hydrogen Balmer decrement, and \hbetaradio{} with \ebv{} \revtext{shown in stars}. Using such a ratio we can compare if the dust correction applied to the observed \hbeta{} line is completely accounted for by the Balmer decrement. For the cases when this ratio is below 1, we can interpret this as evidence of a fraction of ionizing radiation emitted by the galaxy being missed in the Balmer lines. Note that for a few sources the uncertainties of \ebv{} are too small and are not apparent on the figure.

\revtext{From Figure \ref{fig: hbeta flux comparison} (left-axis), we find that many sources in our sample show evidence of dust attenuation at optical wavelengths (e.g., in \bbten{}, \bbthirteen{} and J0159+0751). It is intriguing to find the presence of dust in such low-mass and low-metallicity systems. For J1032+4919 and J1355+4651 this ratio is consistent with 1 within the uncertainities thus suggesting a lack of dust attenuation in these systems. We notice from the right-axis in Figure \ref{fig: hbeta flux comparison} that after dust-correction from the Balmer decrement to the \hbeta{} line, the flux density is consistent with \hbetaradio{} for these sources. This suggests that there is no ionizing radiation being missed in the Balmer lines. For \bbten{} despite the dust-correction from the Balmer decrement to the \hbeta{} line, it is possible that some ionizing radiation emitted by the galaxy is missing from the Balmer lines. However, \bbten{} has an extended structure, thus we could be missing some \hbeta{} flux from the SDSS spectrum. This can reduce the fraction of missing ionizing emission in the Balmer lines than our estimate. J0159+0751 has the highest \ebv{} value in our sample and the ratio between the observed \hbeta{} and \hbetaradio{} flux is also relatively low. Notably, after dust-correcting the \hbeta{} line, the flux density is higher than \hbetaradio{}, however this discrepancy is only weakly significant within the current uncertainties. Future high-sensitivity and high-frequency observations will be essential to unambiguously confirm the presence of dust in these systems.}

\revtext{Overall, by comparing the independently derived thermal flux density using radio data alone with the dust uncorrected and corrected \hbeta{} flux density we find evidence for the presence of dust in a few \xsfg{s} despite being very low-mass and extremely metal-poor systems.}





\section{Discussion}
\label{sec: discussion}

\subsection{RC suppression in xSFGs at low frequencies}

\label{sec: non-thermal suppression discussion}

\begin{figure}
  \resizebox{\hsize}
   {!}{\includegraphics{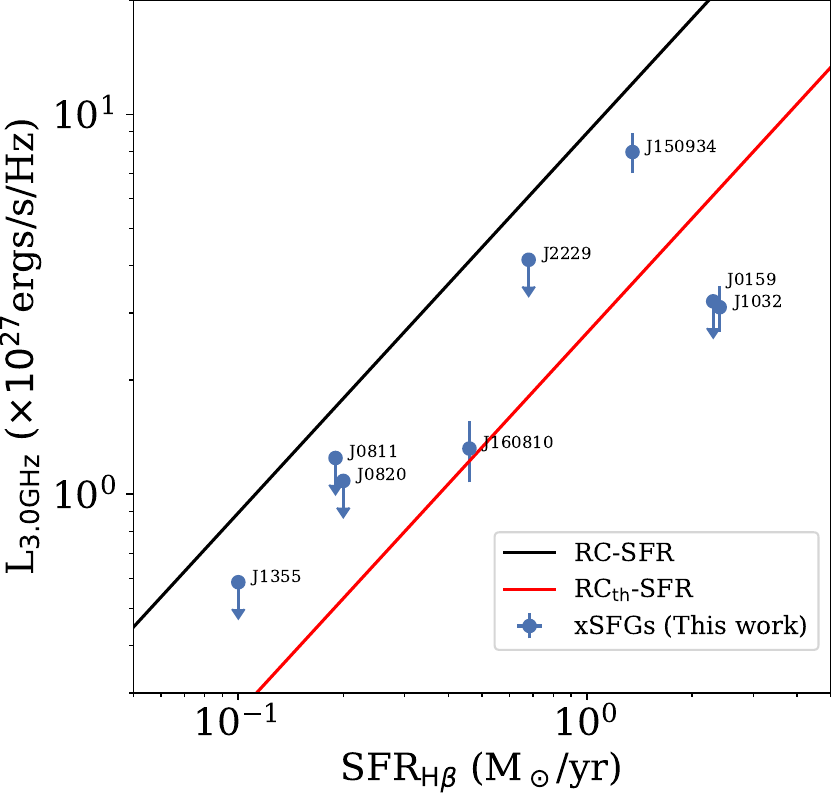}}
  \caption{\xsfg{s} on the RC-SFR relation at $3$ GHz. We show the standard total RC-SFR relation \citep{murphy2011} using a $\alphanth{} = -0.8$ with a solid black line. We also show the RC$_\mathrm{th}$-SFR relation with a red solid line. In both these cases we have used a fixed value of T$_e = 20000$ K. Our \xsfg{s} show a significant suppression of RC emission than that expected from the total RC-SFR relation. See Section \ref{sec: non-thermal suppression discussion} for a discussion.
}
  \label{fig: RC suppression}
\end{figure}

In Figure \ref{fig: RC suppression}, we plot the observed RC luminosity at $3$ GHz against the SFRs for our sample of \xsfg{s}. Our SFRs are taken from the literature (see Table \ref{table: basic properties}) and are derived using the dust-corrected \hbeta{} and H$\alpha$ line luminosity following the \citep{Kennicutt98, Kennicutt12} relation. For reference, we overplot the RC (total)-SFR relation at $3$ GHz for star-forming galaxies from \citet[][Eq. 15]{murphy2011} assuming a \alphanth{} of -0.8, shown in solid black line. The thermal RC (\RCth{}) - SFR relation from \citet[][Eq. 11]{murphy2011} is shown in solid red line. Note that for \bbthirteen{}, since we were unable to measure the flux density at $3$ GHz (see Table \ref{table: radio fluxes}), we have used the modelled RC flux density at $3$ GHz from Section \ref{sec: radio-sed individual cases} and a noise level of $10~ \mu$Jy typically achieved at S-band for our observations. We note that all our \xsfg{s} show suppressed RC emission than that expected from the RC-SFR relation at $3$ GHz, with a few sources even below the \RCth{} - SFR relation. We notice a similar trend in our data at $1.25$ GHz using the \gmrt{} data.

Previous single band radio studies of \xsfg{s} at low-frequencies ($150$ MHz and $\sim1.5$ GHz) had also found that the RC emission in such galaxies is suppressed \citep{chakraborti2012, Sebastian19, Borkar24-BB-LOFAR}, which was indirectly associated with a lack of non-thermal emission. Our finding of a thermally dominant \rcsed{} of xSFGs provides strong evidence for the prevalence of such a scenario. Morevoer the effect of FFA at GHz frequencies could further suppress the RC emission. For example, the RC emission in J1032+4919 was found by \citetalias{Sebastian19} to be highly suppressed at $1.2$ GHz compared to other sources in their sample and was the only undetected source in their study. Our \rcsed{} observation suggests that this non-detection could be due to FFA at $\sim 3.2$ GHz. In some systems RC emission can get suppressed due to FFA even in the presence of  non-thermal emission (e.g., in \bbten{}).

In general, there are several scenarios which can lead to a lack of non-thermal emission: 1) extremely young ages ($< 5$ Myrs) which lack enough SNe \citep[e.g,][]{Hayes07, carilli2008, chakraborti2012, greis2017, Sebastian19}, 2) ionization and bremsstrahlung losses in compact starburst \citep{Murphy09} 3) CR escape \citep{lisenfeld2004,Murphy09, Sebastian19}, 4) strong synchrotron losses due to enhanced magnetic fields \citep{chakraborti2012} and 5) Inverse Compton (IC) losses due to high photon density \citep{condon1992, Murphy09}. The young ages of these galaxies are probably the primary reason for the flat spectrum as based on the observed strong nebular emission lines \citep[e.g., high \ewhbeta{} $>250$\AA,][]{Izotov20-diverseLyA, Izotov24-LyA-metalpoor, Jaskot19, yang2017}. Additionaly, as argued by \citet{Murphy09}, energy loss of CRs due to ionization and bremsstrahlung losses could be an important factor in compact starbursts hosting a dense ISM. The rest of the scenarios involving the lack of CRs (e.g., IC/synchrotron losses, CR escape) would lead to a significant break at high frequencies \citep{klein2018}. We do not observe such a break in the \rcsed{} of \xsfg{s} in our data, however, even higher frequency data are required to completely rule them out. Indeed, \bbten{} is an exception having the highest non-thermal fraction, which possibly has a more complex formation scenario due to a merger with another dwarf galaxy as argued by \citet{Dutta24}.





In summary, \xsfg{s} lack a significant amount of non-thermal emission; however, the exact physical mechanism leading to this suppression \revtext{is} yet to be found.


\subsection{Dependence of LyC escape fraction on radio spectral index}
\label{sec: fesc-alpha}
\begin{figure}
  \resizebox{\hsize}
  {!}{\includegraphics{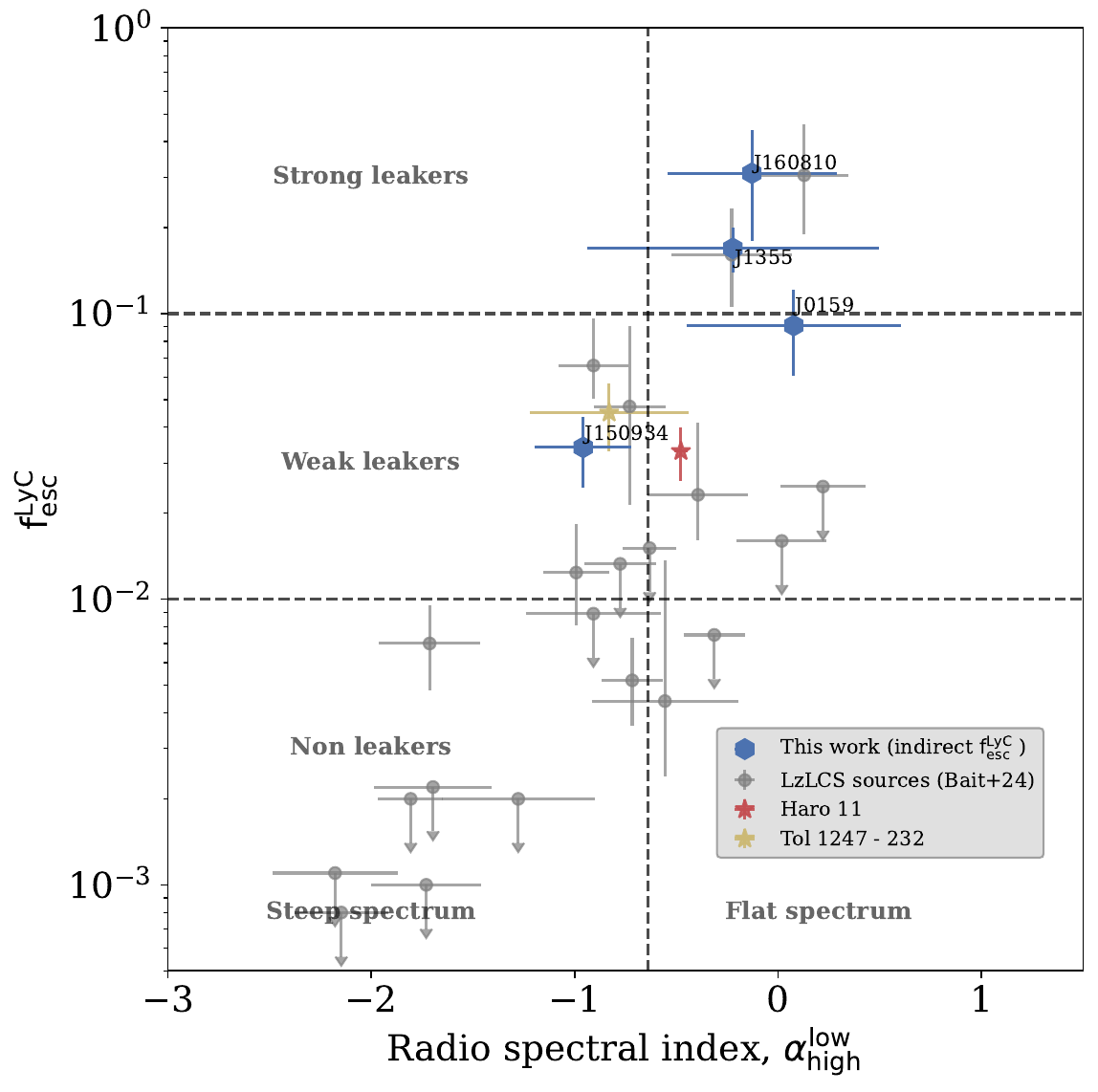}}
  \caption{Dependence of \fesc{} on the radio spectral index, \specindex{}. xSFGs from this paper are shown in blue hexagons. The sub-sample of sources with \vla{} RC observations from \citet{Bait24a} are shown in grey circles, along with Haro 11 (red star) and Tol 1247 - 232 (yellow star). Here the \fesc{} is measured directly using the HST/COS observations and the \specindex{} is measured from observations in the C- and S-bands. For our sample of xSFGs the \fesc{} is estimated indirectly using the empirical relation between \vsep{} and \fesc{} from \citet{Izotov18b-vsep}. The \specindex{} is measured at GHz frequencies which show a detection (see text for details). The two horizontal dashed lines demarcate the regions termed as strong-, weak- and non-leakers. The vertical dashed line represents \alphacs{} $= -0.64$, corresponding to Simple Model, with \fth{} $= 0.1$ and \alphanth{} $= -0.8$. }
  \label{fig: fesc alpha}
\end{figure}

\citet{Bait24a} using RC observations of a sub-sample of LzLCS sources from \citet{Flury22a}, showed that the radio spectral index measured between the \vla{} C- and S-bands (\alphacs{}) correlates with the observed \fesc{}. We refer to this as the \fesc{}-\specindex{} dependence hereon. Here we measure the radio spectral index (\specindex{}) for our detection sample of xSFGs and compare it with the indirectly estimated \fesc{} in Figure \ref{fig: fesc alpha}. The \specindex{} for different sources is measured at several GHz frequencies which show detection. In particular, we use X- and C-bands for J0159+0751 and \bbthirteen{}, Ku- and X-bands for J1355+4651, and C- and S-bands for J150934+373146. The \fesc{} is indirectly estimated using the empirical relation between the \Lya{} velocity peak separation (\vsep{}) and \fesc{} from \citet{Izotov18b-vsep} and are taken directly from \citet{Izotov20-diverseLyA, Izotov24-LyA-metalpoor}. For \bbten{} and \bbthirteen{}, we use the \vsep{} values from \citet{Jaskot19} to estimate the \fesc{}. The \fesc{} values for LzLCS sources are taken from \citet{Flury22a} and \citet{Saldana-Lopez22} which are measured using the ``UV-fit method'' from HST/COS observations. And the \specindex{} is measured between the \vla{} C- and S-bands taken from \citet{Bait24a}. We also overplot in Figure \ref{fig: fesc alpha} two nearby weak leakers, Haro 11 and Tol 1247 - 232 \citep[see ][for references]{Bait24a}. The two horizontal dashed lines split the sources in three broad categories: strong leakers (\fesc{} > 0.1), weak leakers ( 0.01 < \fesc{} < 0.1), and non-leakers (\fesc{} < 0.01). The vertical dashed line represents the \alphacs{} for a simple radio-SED (Simple Model) with \fth{} $= 0.1$ and \alphanth{} $= -0.8$. We generally refer to sources with \specindex{} $> -0.64$ as flat spectrum and \specindex{} $< -0.64$ as steep spectrum. Note that for J1355+4651, the data point exactly falls on another strong leaker source from the LzLCS sample.

Most xSFGs in our sample, except for \bbten{}, are likely strong LyC leakers and possess a flat radio spectrum, as seen in Figure \ref{fig: fesc alpha}. \bbten{} is the only source which has a steeper radio spectrum and is also a likely weak leaker. Taken together we observe that strong leakers are generally described by a flat radio spectrum at GHz frequencies. On the contrary, not all flat spectrum sources show strong LyC leakage, as we observe a few outliers from the LzLCS sample. Moreover, sources with steep spectrum are either weak- or non-leakers. Quite remarkably that our xSFGs with higher \Oratio{} values are less metal enriched (see Figure \ref{fig: alpha physical props} for a comparison between \Oratio{} and metallicity) and also span a lower stellar mass range (\mstar{} $\leq 10^{8}$) than LzLCS sources,  yet they follow the same \fesc{}-\specindex{} dependence. We caution that the \fesc{} for xSFGs is indirectly measured and can be uncertain. Thus, there is tentative evidence that the \fesc{}-\specindex{} dependence holds independently of the stellar mass, metallicity and \Oratio{} ratio for compact star-forming galaxies. Future RC observations particularly focused on strong leakers with direct \fesc{} measurements are required to confirm the universality of the \fesc{}-\specindex{} dependence. 


\citet{Bait24a} argued that the \fesc{}-\specindex{} correlation could be primarily driven by a combination of the age of the starburst and the density variation of the ISM. Since the radio spectral index is a proxy for the age of the starburst \citep{Cannon04, Hirashita06}, the youngest starbursts (ages $< 5$ Myrs) would lack significant non-thermal emission (due to pre-SNe stellar populations) and thus they will have a flat radio spectrum. Additionally, if the source is compact, a large cluster of young stars would be packed in a small region thus boosting the ionized gas density ($N_e \gtrsim 10^3$~cm$^{-3}$). Since the EM is strongly dependent on $N_e$ \footnote{EM $\propto N_e^2$L, where L is the size of the emitting region, usually in the range of 10-100 pc \citep{Hirashita06}.}, this leads to a high EM \citep[in the range of $\sim 10^{7-8}$ pc cm$^{-6}$, see][for a detailed model]{Hirashita06}. Such high EMs are typically a sign of a dominance of young massive star clusters \citep[YMCs; see][for a simulation study]{Inoguchi24_FFA-sims} in the ISM. Thus the radio free-free emission under such conditions would be optically thick at GHz frequencies \citep[e.g.,][]{Johnson03, Hunt04}, further leading to a flat/rising radio spectrum associated with a FFA component \citep[see also ][ for a discussion on this]{ Hirashita06, Bait24a}.

\revtext{Based on our study here, we find evidence for such a thermally dominated \rcsed{} for most of the \xsfg{s} with a couple of systems which are optically thick at GHz frequencies (see Section \ref{sec: radio-sed individual cases}).} This could suggest that strong leakers possess a dominant population of YMCs in their \hii{} regions. YMCs can boost the star-formation efficiency \citep[SFE;][]{Fall10_SFE, Menon24a_bursty_radiation_outlfows} thus leading to a paucity of neutral/ molecular gas thus opening channels for LyC photons to escape \citep{Marques-Chaves24_efficientSFmode}. Such conditions can thus lead to both efficient production and escape of ionizing photons. Overall, this can explain the flat/rising radio spectrum of strong LyC leakers.


For relatively older starbursts ($5 - 10$ Myrs), where the radio free-free emission is optically thin and SNe explosions have accelerated the CR electrons producing a significant amount of non-thermal emission, the radio spectra are steeper. Moreover, in a relatively older starburst pressure-driven expansion of the \hii{} region would lead to a lower EM \citep{Hirashita06}, reducing the effect of flattening of the radio spectra due to FFA. Such conditions in the ISM would in general lead to a weaker LyC leakage, as observed in \bbten{} which has a steep radio spectrum, with a \nuturn{} $< 1$~GHz  and low \fesc{} ($< 10\%$). Finally, \citet{Farcy22} using radiation-magnetohydrodynamics simulations of disc galaxies find that the CR feedback overall leads to a smooth ISM filling it with denser gas. This overall leads to a lower \fesc{} from galaxies. Thus the mere presence (lack) of CRs in weak/non-leakers (strong leakers) could also drive the \fesc{}-\specindex{} dependence and explain its possible existence over a wide stellar mass range ($\log~$\mstar{} $= {6-11}$).  More work is needed on both the observation and modelling front to understand the exact physical mechanism(s) responsible for the \fesc{}-\specindex{} dependence.

\subsection{Relations between \specindex{}, ionization state and metallicity}
\label{sec: alpha ionization}

\begin{figure*}
  \resizebox{\hsize}
  {!}{\includegraphics{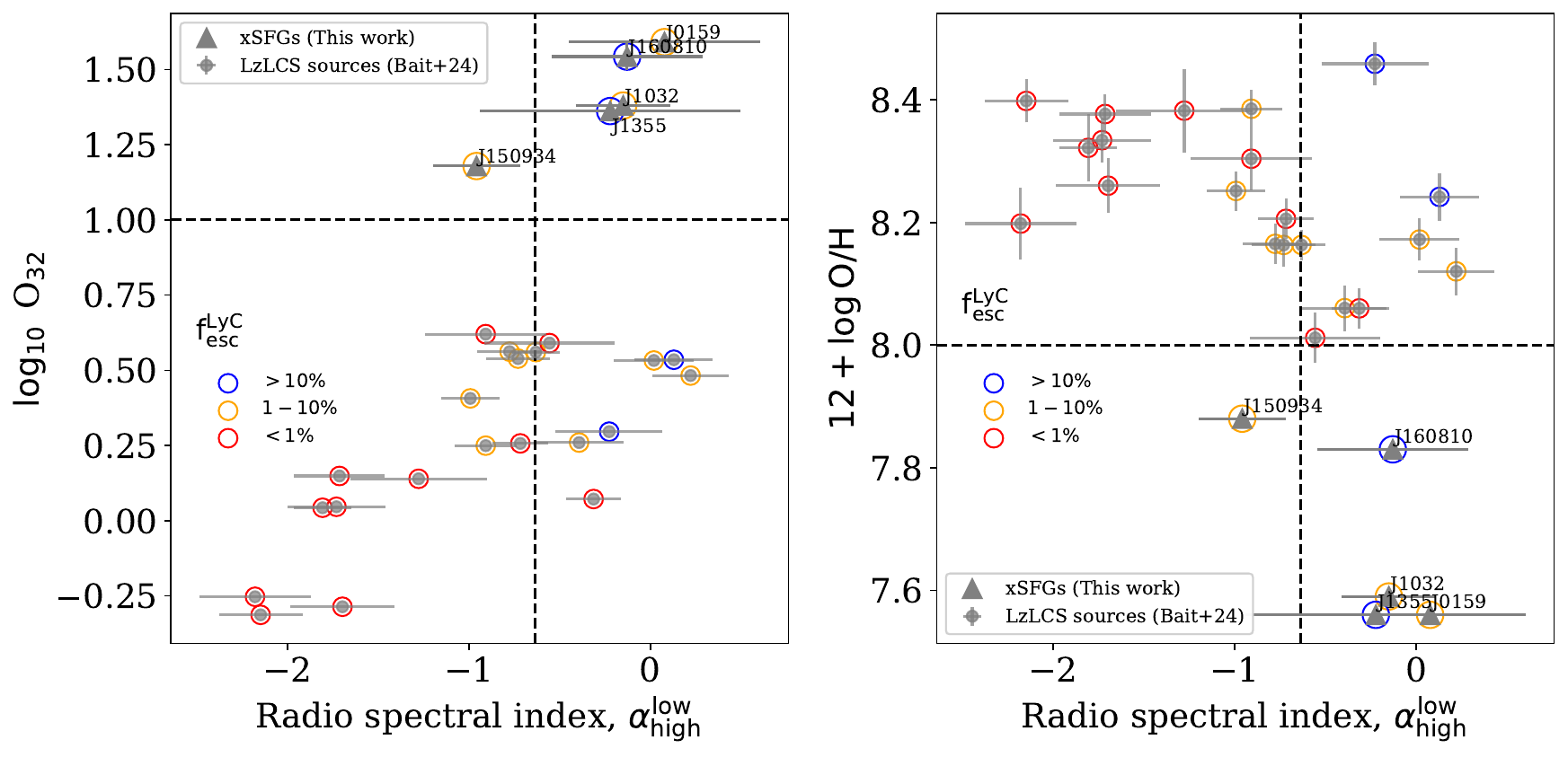}}
  \caption{Relations between \Oratio{} ratio and metallicity ($12+\log\mathrm{O/H}$) with radio spectral index for \xsfg{s} (grey triangles). We show the LzLCS sources with RC measurements from \citet{Bait24a} (grey circles). Around each source we draw circles in blue (\fesc{} $> 10\%$), yellow (\fesc{} $= 1 - 10\%$) and red (\fesc{} $< 1\%$). See Figure \ref{fig: fesc alpha} and text for details on how the \fesc{} values are derived.}
  \label{fig: alpha physical props}
\end{figure*}

High values of \Oratio{} ratio in galaxies  could be associated with low-metallicity, high ionization parameter or the presence of density bounded \hii{} regions \citep{Jaskot-Oey13, Nakajima14-Ionization-Stat}. A high \Oratio{} ratio is found to be a necessary but not a sufficient condition for high \fesc{} in galaxies \citep{Izotov18b-vsep, Jaskot19,  Nakajima20, Flury22b}. Thus we investigate  if the \fesc{}-\specindex{} dependence could be driven by the dependence of \specindex{} on \Oratio{} or metallicity. 

The dependences between \Oratio{}-\specindex{} and metallicity-\specindex{} are presented in the left- and right-panel respectively of Figure \ref{fig: alpha physical props} for xSFGs (in triangles). Such relations were studied by \citet{Bait24a} for a sub-sample of LzLCS sources which are shown in filled circles. This sub-sample lacked sources with very high \Oratio{}, which is covered by our xSFGs sample. We draw coloured circles around these points differentiating them in three categories: strong- (blue circles), weak- (orange circles) and non-leakers (red circles). Note that although J0159+0751 is classified as a weak-leaker, its \fesc{} indirectly derived from the \vsep{} is $\sim 9.1\%$, thus very close to a strong leaker. Overall, we extend the positive correlation between \Oratio{} and \specindex{} found by \citet{Bait24a} for the LzLCS sources, to the highest \Oratio{} ratios ($> 10$). And indeed as previously found the \fesc{} generally increases with increasing \Oratio{} ratio \citep[e.g.,][]{Izotov18b-vsep}. We also observe several outliers to this trend, wherein the flat spectrum sources have a range of \Oratio{} (and  \fesc{} values). The dependence of metallicity on \specindex{} is not so tight. We do observe that the high metallicity sources from LzLCS have steep spectra (and low \fesc{}), although the flat radio spectrum sources are a mixed population. Thus as previously discussed, not all flat spectrum radio sources have high \fesc{} (or high \Oratio{}/low-metallicity). Similar to the \Oratio{} ratio, it appears that a flat radio spectrum is a necessary but not a sufficient condition for strong LyC leakage from galaxies. 

Based on the finding that not all high \Oratio{} ratio sources have high \fesc{}, several studies have suggested that there could be additional physical properties which govern LyC leakage from galaxies, with the viewing angle being a likely possibility \citep[e.g.,][]{Izotov18b-vsep, Jaskot19, Nakajima20, Katz20_orientation, Bassett19_lack_of_O32_fesc}. In our xSFGs, we have such an example of \bbten{}, which has a high \Oratio{} value ($> 10$), and low metallicity comparable for instance to \bbthirteen{}. Yet, their \fesc{} values differ by an order of magnitude. However, we notice that their \specindex{} values vary considerably and can be used to predict their indirect \fesc{} values better (see Figure \ref{fig: fesc alpha}). On the other side, we also have sources from the LzLCS sample with low \Oratio{} that are strong leakers. Here again, such strong leakers have flat \specindex{}. Thus it appears that strong LyC leaking galaxies possess a flatter radio spectrum. Certainly, more data is needed to confirm if \specindex{} could be a primary parameter governing LyC escape in the strongest leakers. If true, this along with the overall correlation between \specindex{} and \fesc{}, and the fact that the \specindex{} is a spatially averaged property could suggest that at least to a first order the orientation might not be a primary driver of \fesc{} \citep[see ][for a more detailed discussion]{Bait24a}. 


In summary, estimating \fesc{} from galaxies is a multi-variate problem \citep[e.g.,][]{Flury22b, Jaskot24-multivariate}, and along with the \Oratio{} ratio, metallicity, dust content and other physical properties, the radio spectral index can play an important role.

\subsection{Overall implications from the \rcsed{} of xSFGs on reionization and high-$z$ galaxies found with JWST}
\label{sec: rcsed implications}

The \rcsed{} of several xSFGs is found to possess a flat spectral index along with evidence for a turnover at GHz frequencies in a few cases, possibly owing to FFA.  These \rcsed{s} are akin to those found in ultra-compact and hyper-compact \hii{} regions in our Galaxy \citep{Wood_UCHII_radio, Yang21_HCHII_radio-sed, Patel24}. The presence of FFA at GHz frequencies is predicted to be a strong observational signature for the presence of YMCs from simulations \citep{Inoguchi24_FFA-sims}. 

Thus our study suggests a picture wherein the extreme star-forming complexes in \xsfg{s} could be dominated by YMC(s) with pre-SNe stellar population. This leads to a high star-formation rate surface density together with a lack of SNe feedback which boosts the star-formation efficiency (SFE) \citep{Fall10_SFE, Menon24a_bursty_radiation_outlfows} and can explain their extreme star-formation properties \citep{Jeckmen23-delayedfeedback, Marques-Chaves24_efficientSFmode}. Such a high SFE can lead to a lack of gas in such galaxies. This can explain the lack of \hi{} detections in low-$z$ green-pea/blueberry galaxies with high \Oratio{} ratio  \citep{Kanekar21_HI-GP, Chandola24b} and the spatial offset between \hi{} and peak of the SF, e.g., in \bbten{} \citep{Dutta24}. Together these effects lead to conditions which are suitable for a significant escape of LyC photons \citep{Marques-Chaves24_efficientSFmode, Menon24a_bursty_radiation_outlfows}.

Since xSFGs are excellent analogues of high-$z$ galaxies \citep[e.g.,][]{Schaerer16, Schaerer22, Rhoads23}, this suggests that the very high-$z$ ($z \geq 6$) galaxies found by JWST might also possess a dominant population of YMCs in their \hii{} regions. Indeed several studies find a dominance of such very dense starclusters in high-$z$ JWST galaxies \citep{Vanzella23, Adamo24, Messa24a, Fujimoto24}. Consequently, these galaxies are expected to be strong LyC leakers and major contributors to reionization.  

Finally, our \rcsed{} study suggests that future deep and ultra-deep radio surveys targeting high-$z$ JWST galaxies with the SKA and ngVLA need to be sensitive to the free-free emission from these galaxies as they are expected to lack a dominant fraction of non-thermal emission. 



\section{Conclusions}
\label{sec: conclusions}

We present new multi-frequency radio continuum observations of $8$ low-$z$ \xsfg{s} using the \gmrt{}, \vla{} and archival \lofar{} telescope to study their \rcsed{}. \xsfg{s} are a population of rare metal-poor ($ 12+\log\mathrm{O/H} < 8.0$), compact starburst galaxies at low-$z$. Using various indicators they are found to be excellent analogues of high-$z$ ($z > 6$) galaxies found with JWST. We  studied \revtext{and modelled} the \rcsed{} in a wide frequency range from $\sim$1 to 15 GHz (with a lower limit of 150 MHz for a few sources) \revtext{using a Bayesian analysis,} to gain novel insights on the nature of extreme star-formation properties and the escape of LyC from these galaxies. 

\revtext{We perform a Bayesian modelling of five \xsfg{s} which shows detections in one or more bands, thus allowing us to put useful constraints on the physical parameters. We also incorporate non-detections in the rest of the bands in our Bayesian analysis. Further, to better constrain the parameters, we also show results using priors on the thermal component (\sth{}) using information from the dust-corrected \hbeta{} flux density.}. Our main results can be summarized as follows: 
   \begin{enumerate}
      \item \revtext{The \rcsed{} of three \xsfg{s} (\bbthirteen{}, J0159+0751 and J1355+4651) shows a flat radio spectrum between $6-15$ GHz. We model these sources using a \revtexts{SFG-thin} model composed of a thermal and non-thermal components. Our Bayesian analysis provides evidence for a thermally dominant \rcsed{} with a \fth{} in the range of $\revtexts{0.5 - 0.7}$. These results remain consistent even when using well-informed priors on \sth{} from the dust-corrected \hbeta{} line, thus providing further support for the thermally dominant scenario.} 
      
      \item \revtext{In the case of J1032+4919, we model the \rcsed{} using a combined \revtexts{SFG-thick} framework, jointly fitting bands with both detections and non-detections. Our Bayesian analysis favors a \rcsed{} characterized by a high thermal fraction (\fth{} $\sim \revtexts{0.5}$), together with a FFA component corresponding to a large emission measure (EM $\sim (9 - 10) \times 10^{7}$ pc cm$^{-6}$).}

      \item \revtext{\bbten{} is the only source in our sample which shows a steep spectrum at high frequencies ($> 1$ GHz) and a flattening below it. Using a \revtexts{SFG-thick} model, our Bayesian modelling with priors on the \sth{} from the \hbeta{} flux density suggests that it can be well modelled with an \fth{} $\sim 0.2$ and a relatively lower EM of $\sim 8 \times 10^{5}$ pc cm$^{-6}$. }

      \item \revtext{Thermally  dominant radio spectra and/or the presence of a FFA component in such systems can explain RC suppression found in previous studies (Sec. \ref{sec: non-thermal suppression discussion}) and their extreme properties (Sec. \ref{sec: rcsed implications}}).
      

      \item Using indirect estimates of \fesc{} in \xsfg{s}, we confirmed the relation between \alphacs{} and \fesc{} from \citet{Bait24a}, particularly for strong leakers (Figure. \ref{fig: fesc alpha}).
      
      \item We also extended the previously found empirical relation between \alphacs{} and \Oratio{} ratio to high values of \Oratio{} (Figure \ref{fig: alpha physical props}).
   \end{enumerate}

\revtext{Future deep low-frequency observations will be crucial to constrain the non-thermal spectral index (\alphanth{}) and the amount of sub-dominant non-thermal component in these systems to better understand the role of SNe feedback in driving the extreme properties of these sources.}

\section*{Acknowledgements}

We would like to thank the referee for their useful comments. This work was supported by the National Science Foundation under Cooperative Agreement 2421782 and the Simons Foundation grant MPS-AI-00010515 awarded to the NSF-Simons AI Institute for Cosmic Origins — CosmicAI, https://www.cosmicai.org/. O. Bait (OB) was also supported by the {\em AstroSignals} Sinergia Project funded by the Swiss National Science Foundation where part of the work was completed. OB would like to thank Heidi Medlin for help with the VLA scheduling.  We thank the staff of the GMRT who have made these observations possible. GMRT is run by the National Center for Radio Astrophysics (NCRA) of the Tata Institute of Fundamental Research. OB would also like to thank NCRA for hosting him where part of this work was completed. Y.I. acknowledges support from the National Academy of Sciences of Ukraine (Project No. 0126U000353) and from the Simons Foundation.

      LOFAR is the Low Frequency Array designed and constructed by ASTRON. It has observing, data processing, and data storage facilities in several countries, which are owned by various parties (each with their own funding sources), and which are collectively operated by the ILT foundation under a joint scientific policy. The ILT resources have benefited from the following recent major funding sources: CNRS-INSU, Observatoire de Paris and Université d'Orléans, France; BMBF, MIWF-NRW, MPG, Germany; Science Foundation Ireland (SFI), Department of Business, Enterprise and Innovation (DBEI), Ireland; NWO, The Netherlands; The Science and Technology Facilities Council, UK; Ministry of Science and Higher Education, Poland; The Istituto Nazionale di Astrofisica (INAF), Italy.

This research made use of the Dutch national e-infrastructure with support of the SURF Cooperative (e-infra 180169) and the LOFAR e-infra group. The Jülich LOFAR Long Term Archive and the German LOFAR network are both coordinated and operated by the Jülich Supercomputing Centre (JSC), and computing resources on the supercomputer JUWELS at JSC were provided by the Gauss Centre for Supercomputing e.V. (grant CHTB00) through the John von Neumann Institute for Computing (NIC).

This research made use of the University of Hertfordshire high-performance computing facility and the LOFAR-UK computing facility located at the University of Hertfordshire and supported by STFC [ST/P000096/1], and of the Italian LOFAR IT computing infrastructure supported and operated by INAF, and by the Physics Department of Turin university (under an agreement with Consorzio Interuniversitario per la Fisica Spaziale) at the C3S Supercomputing Centre, Italy.

\section*{Data Availability}
Raw \vla{} data used in this work are available online in the \vla{} archive (\href{https://data.nrao.edu/portal/}{https://data.nrao.edu/portal/}) under project code: 23A-138 \& 24A-103. Raw \gmrt{} data are available in the GMRT Online Archive (\href{http://naps.ncra.tifr.res.in/goa/}{http://naps.ncra.tifr.res.in/goa/}) under proposal ID: 43\_062 and 45\_075. \lofar{} LoTSS DR2 data used in this work is publicly available at \href{https://lofar-surveys.org/dr2_release.html}{https://lofar-surveys.org/dr2\_release.html}.



\bibliographystyle{mnras}
\bibliography{references.bib} 




\appendix

\section{Deriving the radio-based \hbetaradio{} flux density} \label{appendix sec: radio-based Hbeta flux details}

We derive the expected \hbeta{} line flux density from radio (\hbetaradio{}) following \citetalias{Hunt04} (see their Appendix Eq. A10) which relates the radio-derived thermal flux density and the \hbeta{} line flux density. This relation has been found to be better applicable for low-metallicity \hii{} with high electron temperatures ($> 10,000$ K) which is the case for \xsfg{s}. Here we also assumed a ionized Helium to Hydrogen ratio of 0.08 which is typical of low-metallicty environments \citep{Melnick92, Hunt04}. We used the T$_\mathrm{e}$ values from \citet{Izotov20-diverseLyA, Izotov24-LyA-metalpoor}. For \bbten{} and \bbthirteen{} we used a fixed value of $20,000$ K. \revtext{We have used the \sth{} derived from our Bayesian analysis to estimate the \hbetaradio{}. Specifically we use the values derived from assuming \revtexts{uniform priors} which allows us to independently estimate \sth{} using our radio observations alone.}




\bsp	
\label{lastpage}
\end{document}